\documentclass[%
 reprint, 
 amsmath,amssymb,
 aps, physrev,
]{revtex4-2}

\usepackage{graphicx}
\usepackage{dcolumn}
\usepackage{bm}
\usepackage[dvipsnames]{xcolor}
\usepackage[normalem]{ulem}

\begin{document}

\preprint{APS/123-QED}

\title{\textbf{Toward testing gravity with LSST using $E_G$} 
}%

\author{C. D. Leonard$^{1}$}
\email{danielle.leonard@newcastle.ac.uk}
\author{S. Alam$^{2}$}
\author{R. Mandelbaum$^{3}$}
\author{M. M. Rau$^{1}$}%
\author{S. Singh$^{3,4}$}
\author{C. M. A. Zanoletti$^{1}$}


\affiliation{
$^{1}$School of Mathematics, Statistics and Physics, Newcastle University, Newcastle upon Tyne, NE1 7RU, United Kingdom 
}%
\affiliation{
$^{2}$Tata Institute of Fundamental Research, Homi Bhabha Road, Mumbai 400005, India 
}
\affiliation{
$^{3}$McWilliams Center for Cosmology and Astrophysics, Department of Physics, Carnegie Mellon University, Pittsburgh, PA 15213, USA
\\
$^{4}$Berkeley Center for Cosmological Physics, University of California, Berkeley, CA 94720, USA
}

\collaboration{for the LSST Dark Energy Science Collaboration}

\date{\today}

\begin{abstract}
$E_G$ is a summary statistic that combines cosmological observables to achieve a test of gravity that is relatively model-independent. 
Here, we consider the power of a measurement of $E_G$ using galaxy-galaxy lensing and galaxy clustering with sources from the Rubin Observatory's Legacy Survey of Space and Time (LSST), and lenses from the Dark Energy Spectroscopic Instrument (DESI). We first update the theoretical framework for the covariance of $E_G$ to accommodate this Stage IV scenario. We then demonstrate that $E_G$ offers in principle a model-agnostic test of gravity using only linear-scale information, with the caveat that a careful treatment of galaxy bias is required. We finally address the persistent issue of $E_G$'s theoretical dependence on the measured value of $\Omega_{\rm M}^0$. We propose a framework that takes advantage of the posterior predictive test to consistently incorporate uncertainty on $\Omega_{\rm M}^0$ in tests of gravity with $E_G$, which should be of general use beyond the LSST+DESI scenario. Our forecasting study using this method shows that the prior information available for $\Omega_{\rm M}^0$ is instrumental in determining the power of $E_G$ in the LSST+DESI context. For the full survey dataset, with priors on $\Omega_{\rm M}^0$ from existing CMB data, we find that for some modified gravity scenarios considered, we are likely to be able to reject the GR null hypothesis.
\end{abstract}

\maketitle


\section{Introduction and Motivation}
\label{sec:intro}
\noindent

Current and future optical and infrared photometric and spectroscopic surveys provide a powerful opportunity to test the nature of gravity on cosmological scales. Stage III photometric surveys such as the Kilo Degree Survey (KiDS) \cite{wright2024fifth}, the Dark Energy Survey (DES) \cite{DESDR2} and Hyper Suprime Cam (HSC) \cite{aihara2022third} have begun to explore the potential of weak lensing and galaxy clustering to place constraints on gravity \cite{Joudaki2017, blake2020testing, abbott2023dark}, and the volume of such data will increase by an order of magnitude with the Rubin Observatory's Legacy Survey of Space and Time (LSST) \cite{Ivezic2019}, as well as space-based Stage IV surveys Euclid \cite{Euclid2022}, and the Nancy Grace Roman Space Telescope \cite{Akeson2019}.

A number of different approaches are available when testing gravity using cosmological data. On one hand, we can seek to constrain the parameters of specific individual theories of interest via Bayesian parameter inference methods (see e.g. \cite{joudaki2022gravity, liu2021constraints, piga2023constraints} and many more). This approach is highly valuable in that it promises a clear physical interpretation of the resulting parameter constraints. However, the downside of this approach is that there are many alternative theories of interest. Developing analysis tools and running analyses for each individual theory is a challenging and time-consuming process; covering the entire theory space in this manner is not likely to be feasible. We are therefore also interested in parallel approaches to testing gravity that seek to be model-agnostic.

One model-agnostic approach in common use is to consider a general parametrization which captures a substantial subset of viable alternative gravity theories on linear scales (e.g. \cite{Silvestri2013, bellini2014maximal}). We then place constraints on the parameters that quantify deviations from GR. This approach offers the benefit of constraining relatively general deviations from GR in a single analysis. At the same time, these efficiency gains come at the expense of the specificity and physical interpretability of testing individual theories. Notably, these model-agnostic parametrizations can typically only claim to describe a substantial subset of modified gravity theories in the linear regime of structure growth. In Stage III work, this issue has largely been addressed by imposing scale cuts such that only scales that can be modelled as linear are included in the analysis. However, the incredible statistical power of LSST will severely limit the scales which we will be able to model using linear theory only without risking biases to our parameter inference. Methods to address this issue are under active development \citep{thomas2020cosmological, srinivasan2021cosmological, zanoletti2025principal, tsedrik2024stage}.  

A third option, which is the focus of this work, is to take the approach of performing a consistency check of General Relativity. The $E_G$ statistic \citep{Zhang2007} proposes to do this by strategically combining observables to achieve a measurement which is sensitive to only the Weyl potential and the gravitational potential, with sensitivity to non-gravitational degrees of freedom eliminated.
Since its original proposal, two different practical estimators for $E_G$ have been employed, which differ in whether they probe the Weyl potential via galaxy-galaxy weak lensing \citep{Reyes2010, Blake2015, alam2017testing, amon2018kids+, jullo2019testing, singh2019probing, blake2020testing, rauhut2025testing} or cosmic microwave background (CMB) lensing \citep{Pullen2014, Pullen2015, singh2019probing, zhang2021testing, wenzl2024constraining, wenzl2025atacama}. Alternate proposals to measure $E_G$ via intensity mapping of neutral hydrogen \citep{pourtsidou2016testing, abidi2023model}, using direct measurements of the Weyl potential \citep{grimm2024testing}, and incorporating velocity information via the kinematic Sunyaev-Zel'dovic effect \cite{patki2025probing} also exist in the literature. These various methods share the feature that they aim for $E_G$ to be insensitive to $\sigma_8$ (which describes the amplitude of the matter power spectrum), and to galaxy bias, both of which typically present significant degeneracies with gravitational parameters in other methods.

One issue which has become increasingly relevant to $E_G$ tests of gravity is the fact that the GR prediction for $E_G$ depends on the value of the matter density parameter today, $\Omega_{\rm M}^0$. As cosmological data has become more statistically powerful, we approach a regime where the measurement uncertainty on $E_G$ as estimated from data is comparable to the theoretical uncertainty on the GR prediction for $E_G$ due to uncertainty on our estimates of $\Omega_{\rm M}^0$ (see \cite{amon2018kids+} for a discussion). A consistent statistical framework is required to support the robust interpretation of $E_G$ measurements in the context of this theoretical uncertainty.

In this work, we explore what will be required to make a meaningful measurement of $E_G$ using LSST weak lensing source galaxies together with a spectroscopic galaxy sample from the Dark Energy Spectroscopic Instrument (DESI). 
This paper is structured as follows: in Section \ref{sec:theory}, we review the $E_G$ statistic and its constituent observables as well as other required theoretical background, and we describe the assumed LSST + DESI set-up which we consider. In Section \ref{sec:cov}, we build on existing work in the Stage III context to develop the required covariance modelling framework. In Section \ref{sec:nonlin} we explore the sensitivity of $E_G$ to nonlinear structure formation, including nonlinear galaxy bias, and discuss our method for calibrating the effect of the latter. We then introduce in Section \ref{sec:forecast} the posterior predictive test framework as applied to $E_G$ with uncertain $\Omega_{\rm M}^0$, and determine the power of $E_G$ to test gravity within an LSST+DESI context in several scenarios of true underlying deviations from GR. We discuss and conclude in Section \ref{sec:conc}.

\section{Theory and Setup}
\label{sec:theory}
\noindent
In this Section, we first introduce the $E_G$ statistic and provide theoretical expressions for its constituent observables in its galaxy lensing based formulation. 
We then review the required theoretical basis for the gravity theories we will consider in this work.
Finally, we provide the detailed assumed configuration of the LSST and DESI survey set-ups that we consider.

For the purposes of this Section and the remainder of this work, we use the scalar-perturbed FLRW metric with the following convention:
\begin{equation}
ds^2 = - (1 + 2\Psi) dt^2 + a(t)^2(1 - 2\Phi) \delta_{ij} dx_i dx_j \,.
\label{eq:FLRW}
\end{equation}

\subsection{$E_G$: theoretical background}
\label{subsec:Eg_theory}

The original estimator for the $E_G$ statistic was proposed in \cite{Zhang2007} as:
\begin{equation}
    E_G = \frac{C_{\kappa g}(l, \Delta l)}{3H_0^2a^{-1}\sum_\alpha f_\alpha(l, \Delta l)P_{vg}^\alpha}
    \label{eq:orig_eg}
\end{equation}
where $l$ is the magnitude of the two-dimensional on-sky Fourier-space wave number, $C_{\kappa g}(l, \Delta l)$ is the angular cross-spectrum of weak lensing convergence and galaxy positions in bins of $\Delta l$, $P_{v g}^\alpha$ is the cross-spectrum of velocities and galaxy positions between 3D Fourier wavenumber $k_{\alpha}$ and $k_{\alpha+1}$, and $f_{\alpha}$ is a weighting function which converts $P_{v g}^\alpha$ to an angular power spectrum (not to be confused with $f$, the linear growth rate of structure). The value of the Hubble constant today is denoted by $H_0$ and $a$ is the scale factor as in Eq.~\ref{eq:FLRW}.

The theoretical expectation for $E_G$ is then given by
\begin{equation}
E_G = \left(\frac{\nabla^2(\Psi + \Phi)}{3 H_0^2 a^{-1} f \delta_{\rm M}}\right)_{k=l/\bar{\chi}, \bar{z}}
\label{eq:EG_exp}
\end{equation}
where in this case $f$ is the linear growth rate of structure, $\bar{\chi}$ is the comoving distance corresponding to the weighted average redshift of the lens galaxy sample $\bar{z}$, $\delta_{\rm M}$ is the matter density contrast, and $\Phi$ and $\Psi$ are the potentials which feature in the scalar-perturbed metric of Eq.~\ref{eq:FLRW}.

In GR, and assuming scale-independent galaxy bias, the modelling of $E_G$ can be simplified to \citep{Reyes2010}
\begin{equation}
    E_G = \frac{\Omega_{\rm M}^0}{f(z)} \approx \frac{\Omega_{\rm M}^0}{(\Omega_{\rm M}(z))^\gamma}
    \label{eq:eg_theory_GR}
\end{equation}
where $\gamma=0.55$, $\Omega_{\rm M}(z)$ is the fractional energy density of matter, and $\Omega_{\rm M}^0$ is its value today. We can approximate $\Omega_{\rm M}(z)$ in a flat Universe at late times as
\begin{equation}
    \Omega_{\rm M}(z) = \frac{\Omega_{\rm M}^0(1+z)^{3}}{\Omega_{\rm M}^0(1+z)^3 + (1-\Omega_{\rm M}^0)}.
    \label{eq:OmMz}
\end{equation}
In GR, $E_G$ is thus expected to be a constant with respect to angular or projected radial separation, given by Eq.~\ref{eq:eg_theory_GR}.

In seeking to practically apply $E_G$ using galaxy-galaxy lensing and galaxy clustering data, \cite{Reyes2010} introduces an alternate estimator for $E_G$ in real space:
\begin{equation}
E_G(r_p) = \frac{\Upsilon_{gm}(r_p)}{\beta \Upsilon_{gg}(r_p)}
\label{egreyes}
\end{equation}
where $r_p$ is projected radial separation, and $\beta$ is an observable quantity to be extracted from redshift space clustering measurements in spectroscopic galaxy surveys (see e.g. \cite{percival2011redshift}). From a modelling perspective, it is given by the ratio of the linear growth rate of structure $f$ (at the redshift of the lens galaxy sample) to the linear galaxy bias (of the lens galaxy sample).
$\Upsilon_{gm}(r_p)$ and $\Upsilon_{gg}(r_p)$ are annular differential surface density observables for galaxy-galaxy lensing and projected galaxy clustering, introduced in \cite{Baldauf2010}. 

We now present in detail how we will compute these quantities in this work from a modelling perspective. We begin with $\Upsilon_{gm}(r_p)$, which is defined by convention as in \cite{Baldauf2010} as:
\begin{align}
\Upsilon_{gm}(r_p) &= \Delta \Sigma_{gm}(r_p) - \left(\frac{r_p^0}{r_p}\right)^2 \Delta \Sigma_{gm}(r_p^0)
\label{upgm}
\end{align}
where $r_p^0$ is selected to remove information from the smallest, most theoretically uncertain scales, and $\Delta \Sigma_{gm}(r_p)$ is the differential surface mass density at projected radius $r_p$. It is defined as
\begin{equation}
\Delta \Sigma_{gm}(r_p) = \bar{\Sigma}_{gm}(<r_p) - \Sigma_{gm}(r_p)
\label{DeltaSigma_literal}
\end{equation}
where $\Sigma_{gm}(r_p)$ is the projected surface mass density and $\bar{\Sigma}_{gm}(<r_p)$ is the same quantity averaged over scales less than or equal to $r_p$.
 
$\Delta \Sigma_{gm}(r_p)$ is related to the tangential shear ($\gamma_t$) of background source galaxies about lens galaxies. Within GR, and in the hypothetical and simplified case where we consider the tangential shear of a single source galaxy at $z_s$ and a single lens galaxy at $z_l$ associated with a surface mass distribution at the same redshift, we have: 
\begin{equation}
\Delta \Sigma_{gm}(r_p, z_l) = \gamma_t(r_p, z_l, z_s) \Sigma_{\rm c}(z_l, z_s)
\label{deltasigma}
\end{equation}
where $\Sigma_{\rm c}$ is the the so-called critical surface density. It is defined via 
\begin{align}
\Sigma_{\rm c}^{-1} = \frac{4\pi G}{c^2} \frac{(\chi_s - \chi_l) (1+z_l)\chi_l}{\chi_s}
\label{sigmac}
\end{align}
when $\chi_s \ge \chi_l$, and 0 otherwise. $\chi_s$ and $\chi_l$ are the comoving distances corresponding to the source and lens redshifts and $c$ is the speed of light in a vacuum.

A typical, simplified estimator for $\Delta \Sigma_{gm}$ for a sample of lens and of source galaxies is given schematically by \citep{Mandelbaum2005}
\begin{equation}
    \Delta \Sigma_{gm} = \frac{\sum_{ls} w_{ls} \gamma_t^{ls} \Sigma_{c,\,ls}}{\sum_{ls} w_{ls}},
    \label{eq:DS_obs}
\end{equation}
estimated in bins in $r_p$, where sums are over lens-source pairs. The lens-source pair weights, $w_{ls}$, are typically chosen to downweight lens and source galaxies close in redshift space (see, e.g., \cite{more2023hyper, dvornik2023kids}):
\begin{equation}
    w_{ls} = \Sigma_{c,\, ls}^{-2}.
    \label{eq:weights}
\end{equation}
This adjusts Eq.~\ref{eq:DS_obs} to
\begin{equation}
    \Delta \Sigma_{gm} = \frac{\sum_{ls} \gamma_t^{ls} \Sigma_{c,\,ls}^{-1}}{\sum_{ls} \Sigma_{c,\,ls}^{-2}}.
    \label{eq:DS_obs_weighted}
\end{equation}
In practice, weights will also include a per-source-galaxy factor related to measurement noise, and may also include terms related to the observational selection function of the lens sample. In this work, we make the simplifying assumption that these terms are uncorrelated with redshift and therefore that we can neglect them; including them would simply involve adding an additional redshift-dependent multiplier to Eq.~\ref{eq:weights}.

We can then write a more theoretically oriented version of Eq.~\ref{eq:DS_obs_weighted}, to allow us to model $\Delta \Sigma_{gm}$ in terms of tangential shear and critical density:
\begin{equation}
\Delta \Sigma_{gm}(r_p) = \frac{1}{\bar{w}}\int dz_l \frac{dN_l}{dz_l} \int dz_s \frac{dN_s}{dz_s} \gamma_t(r_p, z_s, z_l) \Sigma_{c}(z_s, z_l)^{-1}
\label{eq:DS_obs_weights}
\end{equation}
where $dN_s/dz_s$ and $dN_l/dz_l$ are the redshift distributions of the source and lens galaxy samples respectively, and $\bar{w}$ is:
\begin{equation}
\bar{w} = \int dz_l \frac{dN_l}{dz_l} \int dz_s \frac{dN_s}{dz_s} \Sigma_{c}^{-2}(z_s,\, z_l).
\label{eq:wbar}
\end{equation}

Eq.~\ref{eq:DS_obs_weights} requires also an expression for $\gamma_t(r_p, z_s, z_l)$, which is given by (see e.g. \cite{leonard2015testing}):
\begin{align}
    \gamma_t&(r_p, z_s, z_l) = -\frac{3H_0^2 \Omega_{\rm M}^0}{2c^2} \int_0^{\chi_s} d\chi \frac{\chi(\chi_s - \chi)(1+z(\chi))}{\chi_s} \nonumber \\ & \times \frac{d}{d \ln r_p} \int_0^{r_p} \frac{dr_p' \, r_p'}{r_p^2} \xi_{gm}\left(\sqrt{(r_p')^2 + (\chi - \chi_l)^2}, z_l\right)
    \label{eq:gammat_lens}
\end{align}
where $\xi_{gm}$ is the three-dimensional two-point position cross-correlation function between the galaxy and matter fields, and we have invoked the small-angle approximation. 

We then propagate Eq.~\ref{eq:DS_obs_weights} for $\Delta \Sigma_{gm}(r_p)$ into an expression for measured $\Upsilon_{gm}(r_p)$. To do so, we follow the procedure of \cite{leonard2015testing}, defining $\Pi \equiv \chi - \chi_l$ and combining Eqs. \ref{upgm}, \ref{eq:DS_obs_weights}, \ref{eq:wbar} and \ref{eq:gammat_lens}. We find:
\begin{align}
\Upsilon_{gm}&(r_p)=\frac{\rho_{\rm c}^0 \Omega_M^0}{\bar{w}}\frac{4\pi G}{c^2} \int dz_l \frac{dN}{dz_l} \int d\Pi\, (\chi_l(z_l)+\Pi)\nonumber \\ &\times [z(\Pi+\chi_l(z_l))+1]  \int_{z_l+\Pi}^{z_s^{\rm max}} dz_s \frac{dN}{dz_s}  \nonumber \\ &\frac{(\chi_s(z_s)-\chi_l(z_l)-\Pi)}{\chi_s(z_s)}\,\Sigma_{\rm c}^{-1}(z_{s}, z_l)\nonumber \\ \times &\Bigg[\frac{2}{r_p^2}\int_{r_p^0}^{r_p}r_p'dr_p' \xi_{gm}\left(\sqrt{(r_p')^2+\Pi^2}, z_l \right)\nonumber \\ &-\xi_{gm}\left(\sqrt{r_p^2+\Pi^2}, z_l)\right) \nonumber \\ &+\left(\frac{r_p^0}{r_p}\right)^2\xi_{gm}\left(\sqrt{\left(r_p^0\right)^2+\Pi^2}, z_l\right)\Bigg],
\label{eq:upgm_expanded}
\end{align}
where $\rho_{\rm c}^0$ is the critical density today. Note that we have used the identity 
\begin{align}
r_p \frac{d}{dr_p}&\left(\frac{2}{r_p^2}\int_0^{r_p} dr_p'\, r_p' f(r_p') \right) = \nonumber \\ &2 \left(f(r_p) - \frac{2}{r_p^2}\int_0^{r_p} dr_p'\, r_p'. f(r_p') \right).
\label{eq:rp_id}
\end{align} 

The integral over $\Pi$ in Eq.~\ref{eq:upgm_expanded} is technically from $-\chi_l$ to $\chi_\infty - \chi_l$ (where $\chi_\infty$ is the comoving distance at $z=\infty$), however in practice any suitably long integration length to capture the correlation behaviour around the lens sample will produce the correct value. We use a projection extent of $\{-\Pi_{\rm max},\, \Pi_{\rm max}\}$ where $\Pi_{\rm max}$ is the line-of-sight comoving distance between $z=0.4$ and $z=1.0$, corresponding to the extent of our DESI lens sample as described in detail below.

Having now an expression which can be used for theoretically computing $\Upsilon_{gm}(r_p)$, we move on to $\Upsilon_{gg}(r_p)$, which is the equivalent annular differential surface density for projected galaxy clustering. It is written as:
\begin{align}
\Upsilon_{gg}(r_p) &= \rho_{\rm c}^0 \Bigg( \frac{2}{r_p^2} \int_{r_p^0}^{r_p} d r_p^\prime r_p^\prime w_{gg}(r_p^\prime) - w_{gg}(r_p) \nonumber \\ &+ \left(\frac{r_p^0}{r_p}\right)^2 w_{gg}(r_p^0) \Bigg)
\label{upgg}
\end{align}
where the projected galaxy clustering correlation function $w_{gg}(r_p)$ is defined as
\begin{equation}
    w_{gg} = \int d\Pi \xi_{gg}(r_p, \Pi)
    \label{wgg}
\end{equation}
and $\xi_{gg}(r_p, \Pi)$ is the three-dimensional galaxy clustering position-position two-point correlation function. We project it over the line-of-sight here because we can only observe redshift space, where $\xi_{gg}$ is anisotropic, and so write it in terms of the line-of-sight separation $\Pi$ and the projected radial separation $r_p$. The integral over $\Pi$ is theoretically over the full extent of the line of sight. In practice, for the cases considered in this work, we find that integrating over a range of $\Pi = \{-900,\,900\}$ Mpc/h achieves numerical convergence.

For convenience, we define a function similar in character to $\Delta \Sigma_{gm}(r_p)$:
\begin{equation}
\Delta \Sigma_{gg}(r_p) = \rho_{\rm c}^0 \left(\bar{w}_{gg}(< r_p) - w_{gg}(r_p)\right)
\label{DeltaSigma_gg}
\end{equation}
such that 
\begin{equation}
\Upsilon_{gg}(r_p) = \Delta\Sigma_{gg}(r_p) - \left(\frac{r_p^0}{r_p}\right)^2 \Delta \Sigma_{gg}(r_p^0).
\label{upgg_DS}
\end{equation}

Note that in this work, we assume that lens galaxies have negligible redshift uncertainty due to being drawn from a spectroscopic sample.

The final component of $E_G$ as defined in Eq.~\ref{egreyes} is $\beta(z)$, defined as 
\begin{equation}
\beta(z) = \frac{f(z)}{b}
\label{beta}
\end{equation}
where $b$ is the large scale galaxy bias and $f(z)$ is the linear growth rate of structure, defined as $f = \frac{d \ln D(a)}{d\ln a}$ where $D(a)$ is the linear growth factor of structure. In GR, $f(z)$ can be calculated theoretically by solving:
\begin{align}
    \frac{\text{d}^2D(a)}{\text{d}t^2} + 2H\frac{\text{d}D(a)}{\text{d}t} - \frac{3}{2}H^2\Omega_{\rm M}(a) D(a) = 0,
\label{beta_de}
\end{align}
where $t$ is physical time.
 
Note that observationally, $\beta$ is extracted from measurements of the redshift-space two-point correlation function of galaxy positions. Redshift-space effects render this 2-point function anisotropic and imprint information about the growth rate of structure; $\beta$ is then extracted via modelling of the anisotropic 2-point function, accounting for these effects (see \cite{kaiser1987clustering} for the linear-theory case). Although the models used for this purpose may not be fully independent of the theory of gravity (particularly when going beyond linear theory), it has been shown that large-scale signatures of alternative theories of gravity can nevertheless be detected via $\beta$ \cite{hernandez2019large}.

\subsection{Gravitational theories}
\label{subsec:mg}
\noindent
In this work, we forecast the potential of the combination of LSST sources with DESI lenses to test gravity with $E_G$. To achieve this, we consider scenarios where an alternative theory in fact describes the Universe. We use two alternative theories of gravity for this purpose: Hu-Sawicki $f(R)$ gravity and normal-branch Dvali-Gabadadze-Porrati (nDGP) gravity. These are selected due to their common use as `workhorse' example theories in cosmology, representing as they do two different screening mechanisms and having a significant body of theoretical and numerical tools to compute quantities of interest. We now provide a review of the basic elements of each theory required in this work.

\subsubsection{nDGP gravity}
\label{subsubsec:nDGPtheory}

Dvali-Gabadadze-Porrati (DGP) gravity belongs to the (higher dimensional) braneworld class of modified gravity models. In DGP, our universe is modelled as a lower-dimensional brane embedded in 5-dimensional Minkowski space. The DGP action is \citep{Dvali_2000}:
\begin{equation}
    \mathcal{S} = \frac{1}{16\pi G_{(5)}} \int \text{d}^5x \sqrt{-g_{(5)}} R_{(5)} + \int \text{d}^4x \sqrt{-g} \left(\frac{R}{16 \pi G} + \mathcal{L}_m \right)
\end{equation}
where the subscript $(5)$ indicates the 5D counterpart to the usual 4D quantities. The cross-over radius $r_c = \frac{G_{(5)}}{2G}$ is a transition scale between the large scales ($\geq r_c$) where 5D dynamics dominate and the small scales ($\leq r_c$), where gravity looks 4-dimensional. The strength of deviation from GR in this paper will be characterised by the parameter $\Omega_{\rm rc} \equiv \frac{1}{4 H_0^2 r_c^2}$, with GR limit $\Omega_{\rm rc} \rightarrow 0$.

The DGP Hubble parameter, neglecting radiation, is given by \citep{deffayet2001cosmology, Schmidt2009}:
\begin{equation}
    H(a) = \sqrt{\Omega_{\rm M}^0 a^{-3} + \Omega_{rc}} \pm \sqrt{\Omega_{\rm rc}}.
\end{equation}
The $(-)$ solution of the expression above is called the normal branch of DGP (nDGP) - this is the solution of interest in this paper. Compared to its self-accelerating $(+)$ counterpart, it is free from ghost instabilities. However, in order to explain late-time acceleration, an additional dark energy component (and therefore an additional degree of freedom) has to be included in the background modelling. This fine-tuned component depends both on the modified gravity parameter and the cosmological constant:
\begin{equation}
    \Omega_{DE}(a) = \Omega_{\Lambda}^0 - 2\sqrt{\Omega_{\rm rc} (\Omega_{\rm M}^0 a^{-3} + \Omega_{\Lambda}^0)}.
\end{equation}

This model provides a minimal extension to $\Lambda$CDM that displays Vainshtein screening and a scale-independent modification to linear structure growth. In the quasistatic regime, we can find the nDGP growth rate by solving the following equation:
\begin{equation}
    \frac{\text{d}^2D(a)}{\text{d}t^2} + 2H(a)\frac{\text{d}D(a)}{\text{d}t} - \frac{3}{2}H(a)^2\Omega_{\rm M}(a) \mu(a)D(a) = 0,
\label{eq:growth_eq}
\end{equation}
with initial conditions $D(a) = a$ and $\frac{\text{d}D(a)}{\text{d}t} = 1$ during matter domination. Here $G$ denotes Newton’s constant, $H$ is the time-evolving Hubble parameter, and $\mu(a)$ is defined as \citep{Schmidt_2009}:
\begin{equation}
    \label{eq:mu_nDGP}
    \mu(a) = 1 + \frac{1}{3\left[1 + \frac{H(a)}{\Omega_{\rm rc}} \left(1 + \frac{a}{3H(a)} \left(\frac{\text{d}H(a)}{\text{d}a}\right)\right)\right]}.
\end{equation}
Comparing Eq.~\ref{eq:growth_eq} with Eq.~\ref{beta} in the GR case, we can easily see the change induced in the linear growth of structure.

We can find the 3D linear matter power spectrum in nDGP gravity (under the continued assumption that we are in the linear and quasistatic regime) via:
\begin{equation}
    P^{\text{nDGP}}_{\delta, {\rm lin}} (k,a) = \left(\frac{D_{\text{nDGP}}(a)\times D_{\text{GR}}(a_{\text{MD}})}{D_{\text{nDGP}}(a_{\text{MD}})\times D_{\text{GR}}(a)} \right)^2 P^{\text{GR}}_{\delta, {\rm lin}}(k,a)
\label{eq:pk_ndgp_lin}
\end{equation}
where $a_{\text{MD}}$ is a value of the scale factor during matter domination.

To find the 3D nonlinear matter power spectrum in nDGP gravity, we use the \texttt{COLA}-based \texttt{nDGPEmu} \footnote{https://github.com/BartolomeoF/nDGPemu} emulator. Compared to \texttt{MG-AREPO} full N-body simulations, it is 3\% accurate for $k = 5h/\text{Mpc}$ and $0\leq z\leq 2$ \citep{fiorini2023fast}.  

\subsubsection{Hu-Sawicki $f(R)$ gravity}
\label{subsubsec:fRtheory}

The $f(R)$ class of gravity theories introduces an action that is beyond-first order in the Ricci scalar $R$:
\begin{equation}\label{eq:fR_action}
    S_G = \int \text{d}^4x\sqrt{-g}\left[\frac{R+f(R)}{16\pi G}\right]
\end{equation}
The additional function $f(R)$ introduces a light scalar degree of freedom $f_R \equiv \text{d}f/\text{d}R$, which leads to a long-range fifth force. We consider more specifically the Hu-Sawicki model \citep{Hu_2007}, with characteristic function:
\begin{equation}
f(R) = -m^2 \frac{c_1(R/m^2)^n}{c_2(R/m^2)^n+1}
\label{eq:husawicki_orig}
\end{equation}
where $m$ is a mass-scale parameter and $c_1$, $c_2$ and $n$ are dimensionless free parameters. $m$ is taken to be
\begin{equation}
    m^2 = \frac{\Omega_{\rm M}^0 H_0^2}{c^2}.
    \label{eq:fR_m2}
\end{equation}
In the high-curvature limit ($R\gg m^2$), and fixing $\frac{c_1}{c^2} = 6 \frac{\Omega_{\Lambda}^0}{\Omega_{\rm M}^0}$ where $\Omega_{\Lambda}^0$ is the fractional energy density of dark energy today), we can write \cite{saezcasares2023emantis}
\begin{equation}
f(R) = -2\Lambda - \frac{f_{R0}}{n}\frac{R_0^{n+1}}{R^n}
\end{equation}
with $f_{R0} = f_R(R_0)$ and $R_0 = R(z=0)$.

In this paper, we will be looking at this version of the theory, with $n=1$ and with a $\Lambda$CDM background. The strength of deviation from GR is then uniquely characterised by the parameter $f_{R0}$, with GR limit $f_{R0} \rightarrow 0$. Under the quasistatic approximation (QSA), with a similar approach as for nDGP, we can find the growth rate by solving Eq.~\ref{eq:growth_eq}, with:
\begin{equation}\label{eq:mu_ka_fR}
    \mu(k,a) =  1 + \frac{(k/a)^{2}}{3\left((k/a)^2 + \frac{H_0^2(\Omega_{\rm M}^0 a^{-3} +4(1-\Omega_{\rm M}^0))^3}{2f_{R0}(4-3\Omega_{\rm M}^0)^2}\right)}
\end{equation}
and similarly for the 3D linear matter power spectrum under QSA:
\begin{equation}
    P^{f(R)}_{\delta, {\rm lin}} (k,a) = \left(\frac{D_{f(R)}(k,a) \times D_{\text{GR}}(k,a_{\text{MD}})}{D_{f(R)}(k,a_{\text{MD}})\times D_{\text{GR}}(k,a)} \right)^2 P^{\text{GR}}_{\delta, {\rm lin}}(k,a).
\label{eq:pk_fR_lin}
\end{equation}

 We compute the 3D nonlinear matter power spectrum for $f(R)$ gravity by using the \texttt{e-MANTIS} emulator \footnote{https://pypi.org/project/emantis/}. When compared to full N-body simulations, it shows a maximum deviation of 3\% for $k \approx 7h/\text{Mpc}$ and $0 \leq z \leq 2$ \citep{saezcasares2023emantis}.

\subsection{Assumed survey set-up and other fiducial parameters}
\noindent
We now describe the assumed set-up of the forecast LSST+DESI $E_G$ measurements considered in this work. We consider an observational scenario in which the lensing source sample is from LSST (we consider both Year 1 and Year 10 scenarios), and the spectroscopic galaxy clustering and lens sample is the Dark Energy Spectroscopic Instrument I (DESI-I) Luminous Red Galaxy (LRG) sample, where DESI-I is the initial 5-year DESI programme which commenced in 2021. We assume an overlap area of 5000 deg$^2$ for these two samples. This number falls between the `baseline' and the `larger extragalactic footprint' scenarios for LSST observing strategy as discussed in \cite{lochner2022impact}.

\begin{figure*}
\centering
\includegraphics[width=0.45\textwidth]{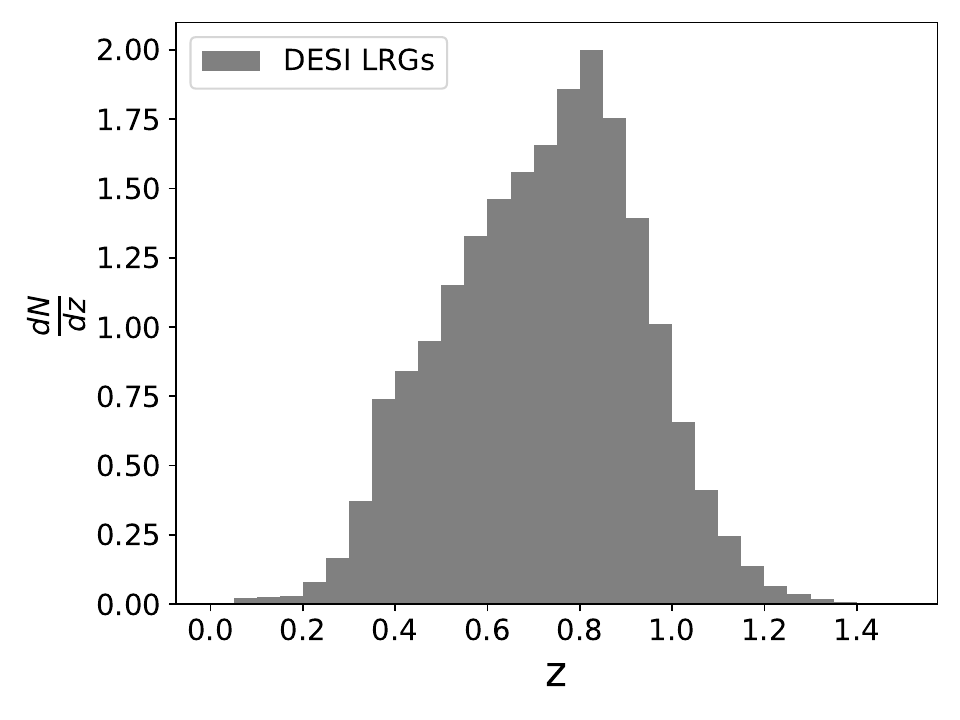}
\includegraphics[width=0.45\textwidth]{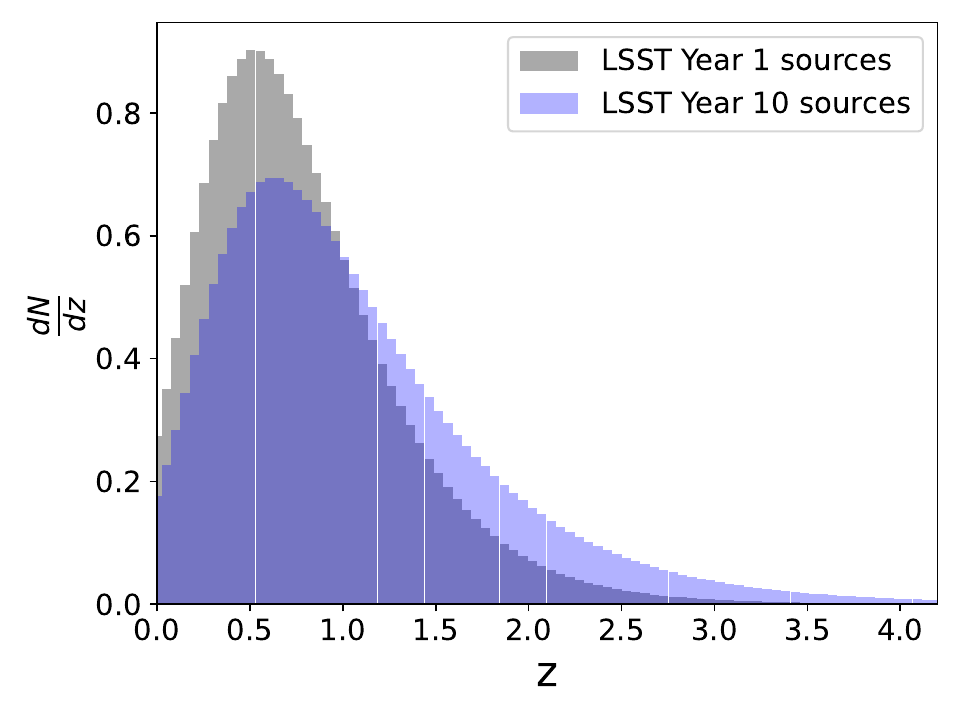}
\caption{{\it Left:} Redshift distribution of DESI LRG sample, data points from \citep{zhou2023target}. Right: Redshift distribution of LSST Year 1 and LSST Year 10 source sample. Numerical values of $
\frac{dN}{dz}$ reflect in all cases the fact that distributions are normalised to unity.}
\label{fig:dNdz}
\end{figure*}

\subsubsection{DESI LRGs}
\label{subsubsec:DESI_specs}
The DESI-I LRG sample is expected to have a comoving volume density of $\bar{n} = 5 \times 10^{-4}$ (Mpc/h)$^{-3}$ over a redshift range of $z=0.4-0.8$ \cite{zhou2023target}. We take the redshift distribution of the DESI-I LRG sample to be as given in \cite{zhou2023target}, which we display in Figure \ref{fig:dNdz}, making the approximation that the above number density holds across the full redshift range of the sample. The mean redshift of this sample is 0.72. As we will see below, for the covariance calculation we must set the line-of-sight extent of the lens sample; we approximate this as being the comoving distance between $z=0.4$ and $z=1.0$ for the DESI LRGs as this is where the majority of the number density falls. It is worth noting that \cite{rauhut2025testing} measures $E_G$ using DESI LRG lenses with sources from Stage III lensing surveys; the DESI sample considered there does not perfectly map onto our case, as it includes DESI Data Release 1 only and has different lens-source sky overlap. Nevertheless we refer the reader to that work for an $E_G$ measurement with DESI LRG lenses.

\subsubsection{LSST source galaxies}

The weak lensing source galaxies are assumed to correspond to the LSST Year 1 or Year 10 source sample as appropriate, defined in the LSST Dark Energy Science Collaboration (DESC) Science Requirements Document \cite{mandelbaum2018lsst}. 

We follow the method of \cite{mandelbaum2018lsst} and model the redshift distribution of the source galaxies by defining a reference distribution and convolving with a model which accounts for photometric redshift uncertainties. Specifically, the reference source galaxy redshift distribution is modelled with an analytic form given by:
\begin{equation}
    \frac{dN}{dz} \propto z^2 \exp\left(\frac{z}{z_0}\right)^\alpha
    \label{eq:smaildndz}
\end{equation}
with $z_0=0.13$ and $\alpha=0.78$ for Year 1, and $z_0=0.11$, $\alpha=0.68$ for Year 10. We model the effect of photometric redshift uncertainty by convolving with a Gaussian distribution with the form:
\begin{align}
    p&(z, \tilde{z}) \propto \exp\left(-\frac{1}{2}\left(\frac{z - (\tilde{z} + \Delta \bar{z})}{\sigma_z(1+z)}\right)^2 \right)
    \label{eq:gaussian_pz}
\end{align}
where $\sigma_z$ and $\Delta \bar{z}$ are parameters of the photo-z uncertainty model, which will in our analysis take fiducial values of $\sigma_z=0.05$ and $\Delta \bar{z}=0$ for both Year 1 and Year 10.

The resulting source galaxy distribution is then given by
\begin{align}
    \frac{dN}{d\tilde{z}}&= \frac{\int dz \frac{dN}{dz}p(z, \tilde{z})}{\int d\tilde{z} \int dz \frac{dN}{dz}p(z, \tilde{z})} \, 
    \label{dNdzp}
\end{align}
and is displayed in Figure \ref{fig:dNdz}.

We assume a surface density of source galaxies of $n_{\rm eff}=10$ galaxies per arcmin$^2$ in Year 1 and $n_{\rm eff}=27$ galaxies per arcmin$^2$ in Year 10. We take the shape-noise parameter $\sigma_{\gamma}=0.26$ for both Year 1 and Year 10.

\subsubsection{Fiducial cosmological and other parameters}
\noindent
We take our fiducial cosmological parameters to match those of the simulation suite which we use for estimating covariances, described in more detail in Section \ref{subsec:sims} below \citep{reid20142}. Explicitly, our fiducial cosmological parameters are: $\{h = 0.69,\, \Omega_{\rm B}^0 =0.022 h^{2} = 0.04621,\, \Omega_{\rm M}^0 = 0.292,\, n_s = 0.965, A_s = 2.115 \times 10^{-9},  \}$. This value of $A_s$ is equivalent to $\sigma_8=0.82$ within $\Lambda$CDM+GR with these other fiducial cosmological parameters.

When we adopt a linear galaxy bias model below, we specifically make use of a single redshift-independent bias parameter for the DESI-LRG lens sample, defined such that $\delta_{\rm g} = b\,\delta_{\rm m}$ where $\delta_{\rm g}$ is the galaxy overdensity and $\delta_{\rm m}$ is the overdensity for all matter. We take a fiducial linear bias value of $b = 2.33$ for the DESI-LRG sample (following \cite{kitanidis2021cross} as discussed further below in Section \ref{sec:nonlin}, together with details of higher-order bias parameters where relevant). This means that the fiducial linear bias differs slightly from the linear bias as computed from the simulations used to estimate the covariance ($b=2.03$). It has been shown that using a covariance matrix computed at fixed parameter values somewhat discrepant from the true underlying parameter values has no significant effect in cosmological analyses \cite{Kodwani2019effect}, thus, we do not expect any significant impact on our results from this mild discrepancy in galaxy bias value.

We follow \cite{Reyes2010, blake2020testing} in fixing $r_p^0 = 1.5$ Mpc/h, although note that \cite{blake2020testing} tested the impact of varying $r_p^0$ for appropriate small separations and found no effect.

\section{Covariance estimation and properties}
\label{sec:cov}

Existing measurements of $E_G$ via galaxy-galaxy lensing take a variety of approaches to covariance matrix estimation, including via mock galaxy catalogues, jackknife sampling, analytic methods, or some combination thereof. 

A common theme in Stage III $E_G$ covariance estimation is the assumption that the three composite probes of $E_G$ are independent of each other \citep{alam2017testing, amon2018kids+, singh2019probing, blake2020testing, rauhut2025testing}. For $\Upsilon_{gm}$ and $\Upsilon_{gg}$, this is argued to be the case because the covariance of $\Upsilon_{gm}$ is dominant (and dominated itself by shape-noise, which is independent of clustering). Additionally, for some measurements, the sky area covered by the source sample is only a small subset of that covered by the lens sample. In the cases of covariance both between $\Upsilon_{gm}$ and $\beta$, and between $\Upsilon_{gg}$ and $\beta$, covariance in the signals is expected to be small and is often neglected \citep{wenzl2024constraining, abidi2023model, pourtsidou2016testing}. This is because $\Upsilon_{gm}(r_p)$ and $\Upsilon_{gg}(r_p)$ are sensitive to modes perpendicular to the line-of-sight, while $\beta$ is sensitive to line-of-sight information as extracted from the redshift-space power spectrum \cite{kaiser1987clustering}.

Although these assumptions are well-founded for the Stage III measurements cited, it is not {\it a priori} obvious that they will hold in the LSST+DESI case. Shape noise in a galaxy-galaxy lensing measurement is suppressed by the number of lens-source galaxy pairs (and similarly for shot noise in a clustering measurement); at the number of galaxies available in LSST and DESI, shape and shot noise may be subdominant to cosmic variance on some scales. Furthermore, the use of the same sample of lens galaxies to measure all three constituent probes may result in some non-negligible covariance between $\Upsilon_{gm}$ and $\beta$, or else between $\Upsilon_{gg}$ and $\beta$, due to the fact that the same galaxy sample is being used within the measurement of all three observables. At the Stage IV level considered here, this may be an important covariance contribution.

In this Section, we present methods for estimating the covariance between constituent probes of $E_G$ based in both analytic methods and simulations, and discuss the extent to which the above assumptions can be carried forward in the Stage-IV context explored here. 

\subsection{Analytic covariance estimation: theoretical background}
\label{subsec:cov_elem}
We first present analytic expressions for the covariance elements which arise between auto- and cross-combinations of $\Upsilon_{gm}(r_p)$ and $\Upsilon_{gg}(r_p)$. 
Although in this work we use the analytic covariances presented here only as a point of comparison for our covariance matrix estimated from simulations, we present the relevant expressions as theoretical background. We make the approximation of neglecting all non-Gaussian contributions from the analytic expressions in this section; it is, however, straightforward to extend the below to incorporate non-Gaussian terms given the ability to compute these for the standard 3$\times$2pt multiprobe analysis case. 

We begin with the covariances of $\Delta \Sigma_{gx}$ quantities which can then be propagated to $\Upsilon_{gx}$ (with $x$ being $m$ or $g$). We follow closely here the work of \cite{Singh2016} and refer the reader also to \cite{blake2020testing} for a similar analytic treatment. 

Considering first the covariance of $\Delta \Sigma_{gm}(r_p)$, we have (see, \cite{Singh2016} in the Gaussian limit)
\begin{align}
&{\rm Cov}\left[\Delta \Sigma_{gm}\left(r_p\right),\Delta \Sigma_{gm}\left(r_p^\prime\right) \right] =\frac{\left(\overline{\Sigma_c^{-2}}\right)^{-1}}{V} \int \frac{k dk}{2\pi}J_2 \left(k r_p\right)  \nonumber \\ & \times J_2\left(k r_p^\prime\right)\Bigg[\left(P_{g\kappa}(k)\right)^2 + \left(P_{gg}(k)+\frac{1}{n_{\rm l}}\right)\Bigg(  P_{\kappa \kappa}(k) + \frac{\sigma_{\gamma}^2}{n_{\rm s}} \Bigg)\Bigg],
\label{DeltaSigmaCov}
\end{align}
where $n_{\rm l}$ is the volume density of lens galaxies, $n_{\rm s}$ is the effective surface density of source galaxies, $\sigma_\gamma$ is the rms per-component shape noise, $V$ is the volume of the lens galaxy sample and $J_2$ is the second-order spherical Bessel function. $\overline{\Sigma_c^{-2}}$ is defined as
\begin{equation}
\overline{\Sigma_c^{-2}} = \int dz_{l} \frac{dN}{dz_{l}} \int dz_{ph} \frac{dN}{dz_{s}} \Sigma_c^{-2}(z_{l}, z_{s}),
\label{wbar}
\end{equation}
 $P_{\kappa \kappa}(k)$ is given by
\begin{align}
P_{\kappa \kappa}(k) = \int d \Pi \Bigg(\int &d z_s \frac{dN}{dz_s} \int dz_l \frac{dN}{dz_l} \nonumber \\ &\times\frac{\rho_{\rm m}^0}{\Sigma_{\rm c}(\chi_s(z_s), \chi_l(z_l)+\Pi)}\Bigg)^2 P_\delta(k),
\label{pkk}
\end{align}
and similarly $\left(P_{g\kappa}(k)\right)^2$ is
\begin{align}
(P_{g \kappa}(k))^2 = \int &d \Pi \Bigg(\int d z_s \frac{dN}{dz_s} \int dz_l \frac{dN}{dz_l} \nonumber \\ &\times \frac{\rho_{\rm m}^0b}{\Sigma_{\rm c}(\chi_s(z_s), \chi_l(z_l)+\Pi)}\Bigg)^2 P_\delta(k)^2,
\label{pgk1}
\end{align}
where $b$ is the galaxy bias, which could be redshift or scale dependent, and $\rho_m^0$ is the value of the matter density today. We note that the projection over $\Pi$ is here over the entire line-of-sight distance relevant to the lensing measurement, reflecting the appropriate window function.

Similarly, for the covariance of $\Delta \Sigma_{gg}(r_p)$, we follow the derivation of \cite{Singh2016} and take the Gaussian limit to obtain
\begin{align}
{\rm Cov}&\left[\Delta \Sigma_{gg}\left(r_p\right),\Delta \Sigma_{gg}\left(r_p^\prime \right) \right] = \frac{2 \Delta \Pi (\rho_{\rm c}^0)^2 }{V}  \int \frac{k dk}{2\pi} J_2\left(k r_p\right) \nonumber \\ & J_2\left(k r_p^\prime \right) \left(P_{gg}(k)+\frac{1}{n_{\rm l}}\right)^2,
\label{DeltaSigmaCov_gg}
\end{align}
where $\Delta \Pi$ is the line-of-sight extent of the lens sample. 

Finally, we once again follow an analogous derivation to those in \cite{Singh2016} to obtain the covariance of $\Delta \Sigma_{gm}$ with $\Delta \Sigma_{gg}$:
\begin{align}
{\rm Cov}&\left[\Delta \Sigma_{gm}\left(r_p\right),\Delta \Sigma_{gg}\left(r_p^\prime \right) \right] = \frac{2 \rho_{\rm c}^0}{V}\left(\overline{\Sigma_c^{-1}}\right)^{-1} \nonumber \\ & \times \int \frac{k dk}{2\pi} J_2\left(k r_p\right)J_2\left(k r_p^\prime\right)\bar{P}_{g\kappa}(k)\left(P_{gg}(k)+\frac{1}{n_{\rm l}}\right)
\label{DeltaSigma_gm_gg_cov}
\end{align}    
where $\bar{P}_{g\kappa}$ differs from $P_{g\kappa}$ in Eq.~\ref{pgk1} above in that it is projected over the window function suitable for one galaxy clustering measurement and one galaxy-galaxy lensing measurement:
\begin{align}
\bar{P}_{g \kappa}(k) = \int_{\Pi_{\rm min}}^{\Pi_{\rm max}} &d \Pi \Bigg(\int  d z_s \frac{dN}{dz_s} \int dz_l \frac{dN}{dz_l} \nonumber \\ & \times \frac{\rho_{\rm m}^0 b}{\Sigma_{\rm c}(\chi_s(z_s), \chi_l(z_l)+\Pi)}\Bigg) P_\delta(k),
\end{align}
where $\Pi_{\rm min}$ and $\Pi_{\rm max}$ are the limits of the lens sample along the line-of-sight. 

Having now Cov[$\Delta \Sigma_{gx}(r_p),\, \Delta \Sigma_{gy}(r_p')$] for each of $x$ and $y$ equal to $g$ and $m$, we return to the definitions of $\Upsilon_{gm}(r_p)$ and $\Upsilon_{gg}(r_p)$ (Eqs. \ref{upgm} and \ref{upgg} respectively) to find:
\begin{align}
{\rm Cov}&(\Upsilon_{gx}(r_p),\Upsilon_{gy}(r_p^\prime)) = {\rm Cov}(\Delta \Sigma_{gx}(r_p),\Delta \Sigma_{gy}(r_p^\prime)) \nonumber \\ &-\left(\frac{r_p^0}{r_p}\right)^2  {\rm Cov}(\Delta \Sigma_{gx}(r_p^0),\Delta \Sigma_{gy}(r_p^\prime))\nonumber \\&-\left(\frac{r_p^0}{r_p^\prime}\right)^2  {\rm Cov}(\Delta \Sigma_{gx}(r_p),\Delta \Sigma_{gy}(r_p^0)) \nonumber \\&+ \left(\frac{r_p^0}{r_p}\right)^2 \left(\frac{r_p^0}{r_p^\prime}\right)^2 {\rm Cov}(\Delta \Sigma_{gx}(r_p^0),\Delta \Sigma_{gy}(r_p^0)),
\label{covupgm}
\end{align}
where $x$ and $y$ are either $g$ or $m$ as appropriate. Explicit expressions for the averaging in $r_p$ bins as required are found in Appendix \ref{app:covmaths}, along with a visualisation of the correlation matrix estimated for this subset of terms using this analytic method for LSST sources and DESI LRG lenses (Figure \ref{fig:joint_cov_corr_ana}).

\subsection{Covariance estimation via simulations}
\label{subsec:sims}

\begin{figure*}
\centering
\includegraphics[width=0.45\textwidth]{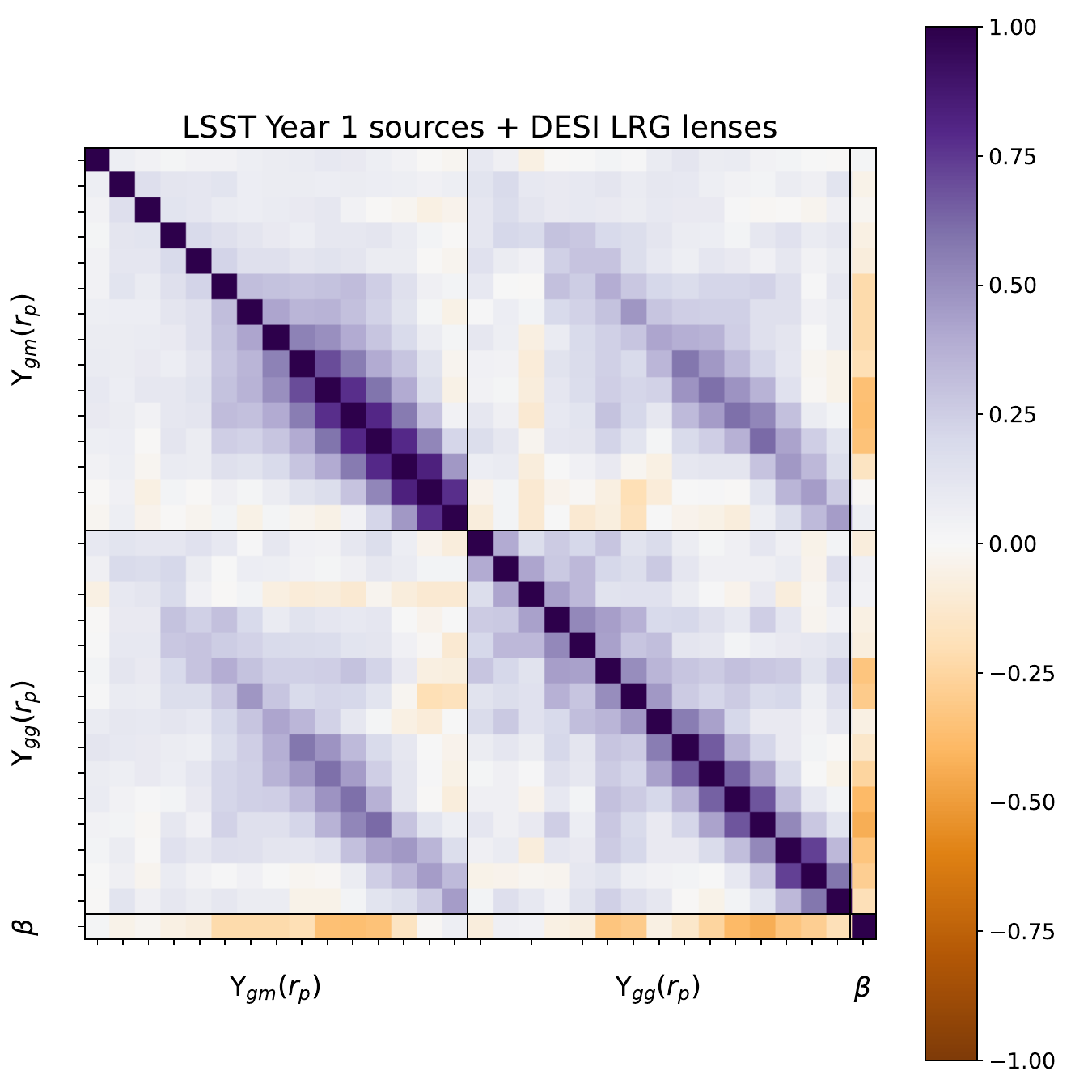}
\includegraphics[width=0.45\textwidth]{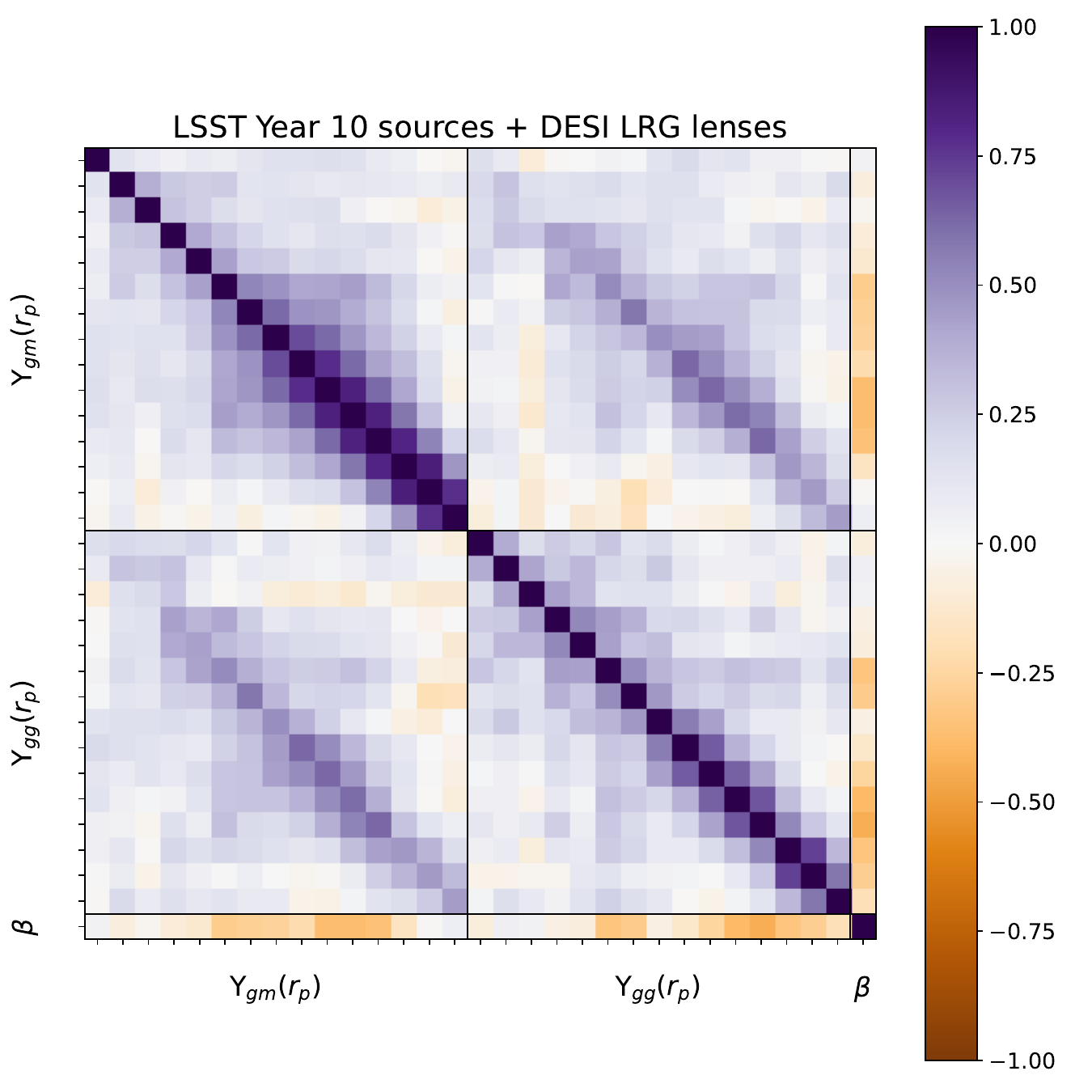}
\caption{Correlation matrices of the constituent probes of $E_G$. Within the $\Upsilon_{gm}(r_p)$ and $\Upsilon_{gg}(r_p)$ sub-blocks, bins increase logarithmically in $r_p$ with bin centres ranging from 1.4 to 87.9 Mpc/h.}
\label{fig:joint_cov_corr}
\end{figure*}

We can also estimate the covariance of the constituent probes of $E_G$ via mock galaxy catalogues based on N-body simulations. In practice, we will use this method in this work. This is primarily due to the fact that the simulation-based method allows us to easily estimate the covariance between $\Upsilon_{gm}$ / $\Upsilon_{gg}$ and $\beta$, for which an analytic formalism is not available in the literature. Where available, we make use of covariance matrices generated analytically as a point of comparison for those estimated from simulations (although we do not expect perfect agreement due to differences in the effects captured in each case).

To obtain the required mock galaxy catalogues, we use the N-body simulations described in \cite{reid20142} (`MedRes'), taking a redshift slice at $z=0.75$ to best match the DESI LRG sample redshift. We run the Rockstar \cite{behroozi2012rockstar} halo finder on this N-body snapshot and populate the resulting haloes with galaxies using a halo occupation distribution (HOD) model fit to the DESI LRG Buzzard simulations between redshift 0.7-0.8 (see \cite{alam2017testing} for the precise form of the HOD model). The result is a galaxy number density of $4.954 \times 10^{-4}$ (Mpc/h)$^{-3}$, in excellent agreement with our assumed DESI LRG number density.

The simulation volume is subdivided into 100 jackknife regions. Given the mock galaxy catalogues, $\Upsilon_{gm}$ and $\Upsilon_{gg}$ are measured in $r_p$ bins as surface mass densities using dark matter particles and galaxies. $\beta$ is obtained by fitting a Convolutional Lagrangian Perturbation Theory model \citep{wang2014analytic} between $30-80$ Mpc/h. We then use standard expressions for jackknife covariance estimation to find a covariance matrix which captures the auto- and cross-variance between all three quantities, rescaling the resulting covariance to account for the difference in the simulation volume vs that expected for the LSST+DESI measurement under consideration. 

We must then correct this matrix to account for the fact that although one can estimate $\Upsilon_{gm}$ easily from the simulations described as a cross-correlation between dark-matter particles and galaxies, this ignores a key feature of the practical observational estimate of $\Upsilon_{gm}$ via galaxy-galaxy lensing: shape noise. Fortunately, shape noise is independent of the other noise terms which are accounted for by the simulation. Therefore, we can construct an independent covariance matrix to capture this effect. We construct this shape-noise covariance matrix by considering only the final term in brackets of Cov[$\Delta \Sigma_{gm}(r_p), \Delta \Sigma_{gm}(r_p')$] (Eq.~\ref{DeltaSigmaCov}), which yields:
\begin{equation}
{\rm Cov}^{\rm SN}[\Delta \Sigma_{gm}(r_p), \Delta \Sigma_{gm}(r_p')] = \frac{\left(\overline{\Sigma_c^{-2}}\right)^{-1}\sigma_\gamma^2}{2\pi V n_l n_s}\frac{\delta^D(r_p - r_p')}{r_p}
\label{eq:sn_cov_DS}
\end{equation}
where $\delta^{\rm D}$ indicates a Dirac delta function.

We propagate this expression to the $\Upsilon_{gm}$ case using Eq.~\ref{covupgm} and average in $r_p$ bins using Eq.~\ref{covupgm_avg} given in Appendix A. Note that in practice, when computing this component, we assume that $\Delta \Sigma_{gx}(r_p^0)$ is measured as $\Delta \Sigma_{gx}$ in the $r_p$ bin containing $r_p^0$. Because this noise component is independent of those captured by our jackknife-estimated covariance matrix, we can simply add the resulting contribution to the $\Upsilon_{gm}(r_p)$ auto-correlation block of the matrix. 

Finally, note that because we use a covariance matrix estimate based on jackknife methods with a finite number of jackknife realisations, when inverting the covariance matrix, we must account for the fact that the inverse of a noisy covariance matrix is not an unbiased estimate for the true inverse covariance. This is accounted for by multiplying the resulting inverse covariance by the so-called  Hartlap factor \citep{hartlap2007your}:
\begin{equation}
    {{\rm Cov}}^{-1} = \frac{n-p-2}{n-1}{{\rm Cov}}^{-1}_*
    \label{hartlap}
\end{equation}
where ${{\rm Cov}}^{-1}_*$ is the uncorrected inverted estimate of the covariance matrix, ${{\rm Cov}}^{-1}$ is the corrected version we will use, $n$ is the number of jackknife realisations, and $p$ is the length of the data vector. 

The full correlation matrices for our \{$\Upsilon_{gm}(r_p)$,\, $\Upsilon_{gg}$,\, $\beta$\} datavector, estimated via the method described in this subsection, are visualised in Figure \ref{fig:joint_cov_corr} for LSST Year 1 and Year 10 sources with DESI lenses. We see that, as we suggested may be the case at the beginning of this Section, the matrix is not fully dominated by the shape noise associated with the source sample. Furthermore, contrary to conventional expectation for Stage III data and as alluded to above \citep{wenzl2024constraining, abidi2023model, pourtsidou2016testing}, we also find non-negligible Cov($\Upsilon_{gm},\, \beta$) and Cov($\Upsilon_{gg},\, \beta$). 

\subsection{$E_G(r_p)$ covariance}
\label{subsec:egcov}

\begin{figure*}
\centering
\includegraphics[width=0.45\textwidth]{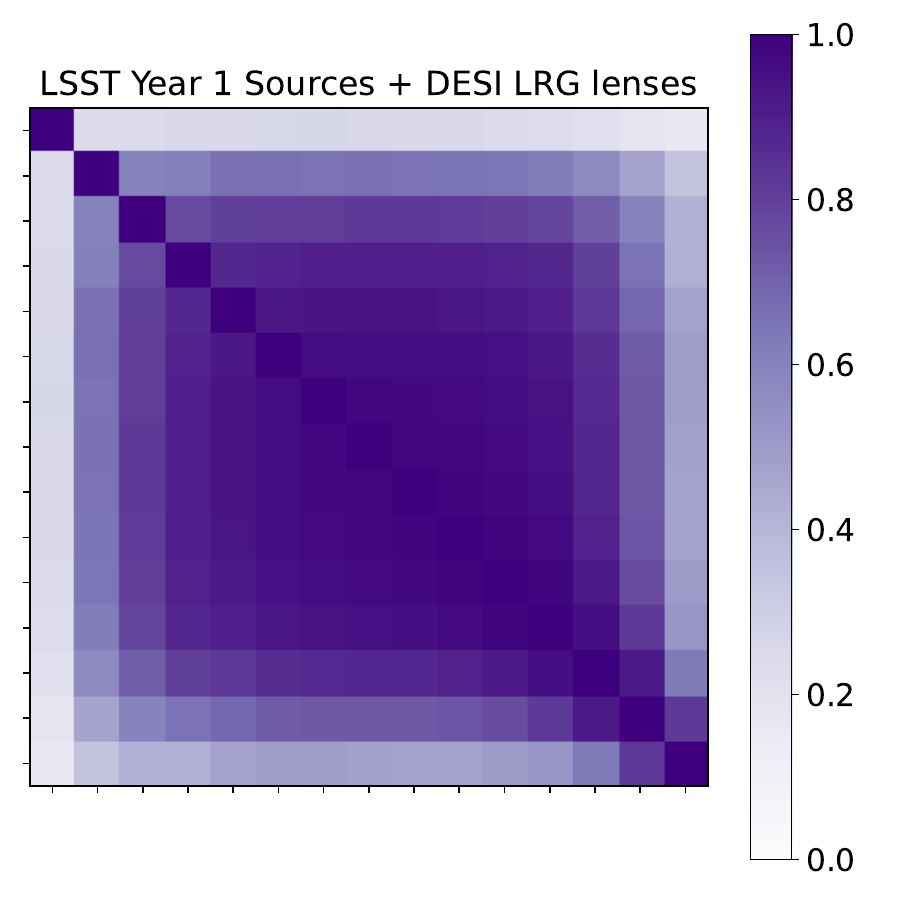}
\includegraphics[width=0.45\textwidth]{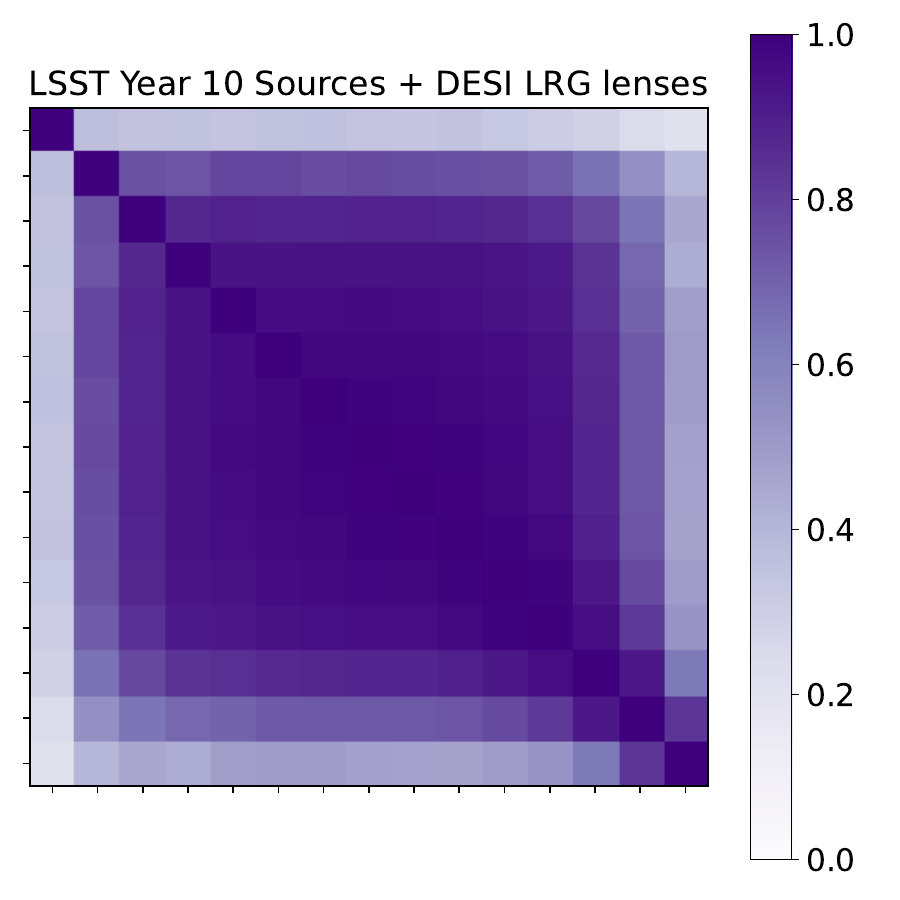}
\caption{Correlation matrices of $E_G(r_p)$. Bins increase logarithmically in $r_p$ with bin centres ranging from 1.4 to 87.9 Mpc/h.}
\label{fig:eg_cov}
\end{figure*}

\noindent
As can be readily seen from Eq.~\ref{eq:EG_exp}, $E_G(r_p)$ is a ratio of noisy quantities. In principle, the ratio of Gaussian-distributed quantities is not necessarily a Gaussian, but under some circumstances can be approximated as such. If the Gaussian approximation holds, we can carry on with the use of a covariance matrix for $E_G(r_p)$ to characterise an approximately Gaussian likelihood. We first therefore verify to what extent the Gaussian approximation holds for our case.

We construct a multi-variate Gaussian using the covariance matrix of the constituent probes and the fiducial data vector. We then jointly draw samples of $\Upsilon_{gm}(r_p)$, $\Upsilon_{gg}(r_p)$, and $\beta$, then use these values to create a population of $E_G(r_p)$ samples. In each $r_p$ bin (call this $r_p^i$), we examine the distribution of $E_G(r_p^i)$ and find that for all bins, the Gaussian approximation is sufficient for our purposes. Note that this diverges from recent work for the case of the CMB lensing $E_G$ statistic, where it was found that $E_G$ is not Gaussian-distributed \citep{wenzl2024constraining}. The validity of the Gaussian likelihood approximation should be verified for each individual data set as it will depend on the specific noise properties in question.

Turning again to our population of $E_G(r_p)$ samples, ${\rm Cov}(E_G(r_p), E_G(r_p^\prime))$ can then be estimated using standard formulae. The correlation matrices for $E_G$ in $r_p$ bins are shown in Figure \ref{fig:eg_cov} for LSST Year 1 and Year 10 sources with DESI LRG lenses. The correlation matrix demonstrates very significant off-diagonal covariance between $r_p$ bins, which is further enhanced in the Year 10 case (where the shape noise is reduced).

\section{$E_G$ and nonlinear scales}
\label{sec:nonlin}

A major issue facing Stage IV weak lensing and galaxy clustering tests of deviations from GR is the challenge of appropriately modelling nonlinear growth (and related effects such as nonlinear galaxy bias) outside of standard gravity. For constraints on parameterised modifications to GR in Stage III surveys, this has often been circumvented by restricting to only linear scales. In Stage IV, we expect this strategy to be highly suboptimal, as increased statistical precision of our measurements will make it much more restrictive (see, e.g. \cite{zanoletti2025principal}). Strategies to overcome this issue are under development \cite{bose2020road, thomas2020cosmological, srinivasan2021cosmological, tsedrik2024stage, zanoletti2025principal}, but these do not map directly onto the $E_G$ case, where the goal is not to constrain parameters but to perform a litmus test for GR.

\begin{figure}
\centering
\includegraphics[width=0.4\textwidth]{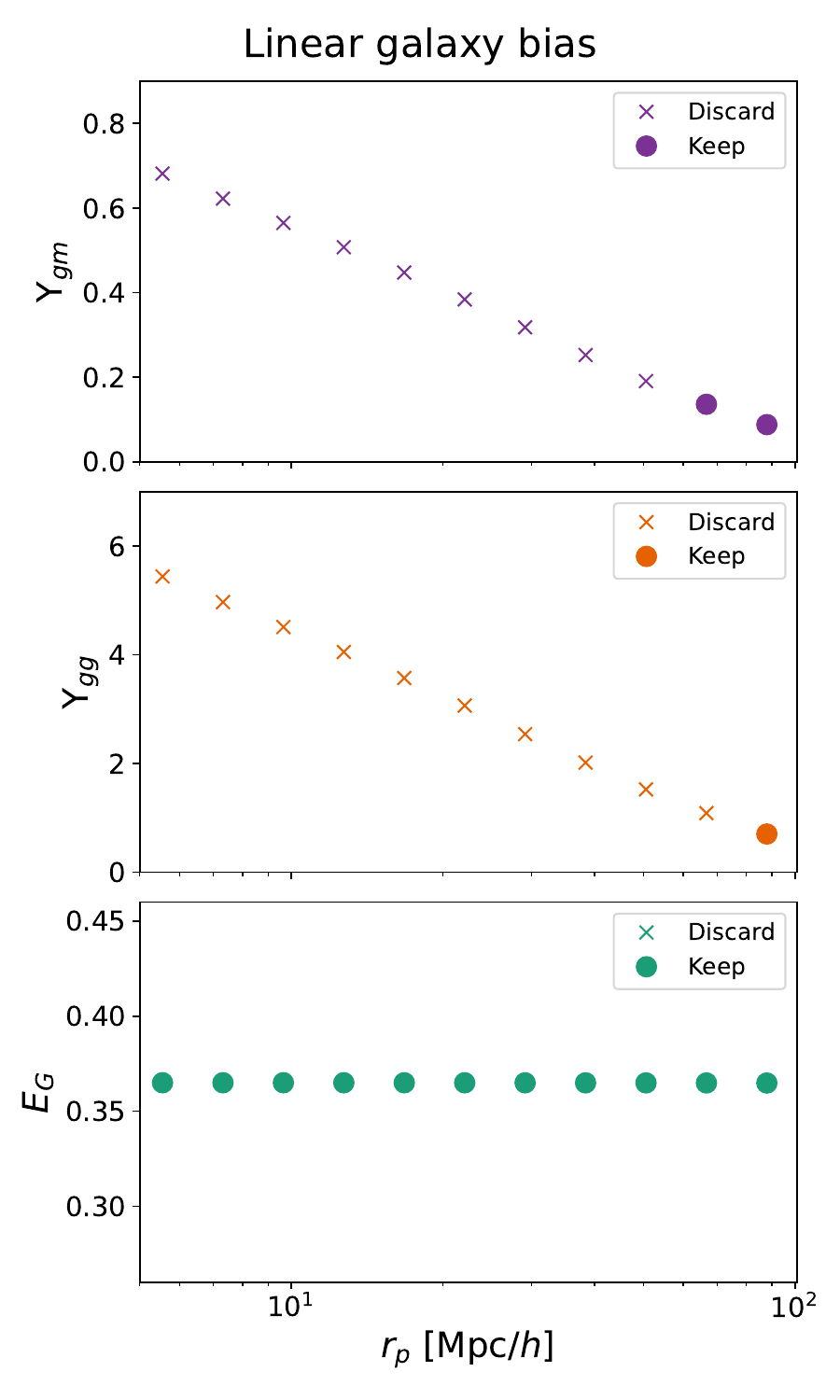}
\caption{Illustration of which points are kept in the linear-only scale cuts scenario for $\Upsilon_{gm}$ (top), $\Upsilon_{gg}$ (middle) and $E_G$ (bottom), assuming a linear galaxy bias. The points plotted are for the fiducial data vector with non-linear modelling from halofit (but linear galaxy bias), with LSST Year 1 sources. The $r_p$ bins kept vs discarded are the same in both the LSST Year 1 and Year 10 source cases.}
\label{fig:scales_to_keep}
\end{figure}

In this Section, we consider the robustness of an LSST+DESI $E_G$  measurement to uncertainties in the modelling of nonlinear growth of structure. We examine the degree to which this depends on the underlying gravity theory and the effect of nonlinear galaxy bias.

We make use of the standard linear criterion (seen elsewhere in the literature in, for example, \cite{DESY1ext, abbott2023dark, ade2016planck, Joudaki2017}) that scales are retained in an analysis for which it is true that:
\begin{align}
\Delta &\chi^2_{{\rm nonlin} - {\rm lin}} = \nonumber \\& \left(\vec{D}_{\rm nonlin} - \vec{D}_{\rm lin}\right) {\rm Cov}^{-1}\left(\vec{D}_{\rm nonlin} - \vec{D}_{\rm lin}\right) \leq 1
\label{chisqreq}
\end{align}
where $\vec{D}_{\rm nonlin}$ is the fiducial theoretical data vector and $\vec{D}_{\rm lin}$ is the fiducial datavector modelled with linear theory only. ${\rm Cov}$ is the data covariance matrix. 

We follow literature convention in imposing the additional constraint that if a given data point is removed by the criteria of Eq.~\ref{chisqreq}, so too should be all data points at smaller scales for the observable in question. This accounts for two main effects. First, at quasilinear scales, the nonlinear behaviour of voids can lead to a `zero crossing' in $\left(\vec{D}_{\rm nonlin} - \vec{D}_{\rm lin}\right)$ \cite{mead2021hmcode}. While this may technically mean that some intermediate-scale data points contribute less to $\Delta \chi^2$ than data points at higher and lower scales, the exact location of the zero-crossing is expected to be sensitive to nonlinear modelling, and so we cut these intermediate points. Second, there may be cases where higher variance and covariance at smaller scales mean that the criteria of Eq.~\ref{chisqreq} would allow us to retain data points at scales which are certainly subject to significant nonlinear effects. Although we could in principle retain these points, we opt to define our criteria such that they are removed to maximise interpretability.

Note also that before applying Eq.~\ref{chisqreq}, following \cite{alam2017testing} and \cite{blake2020testing}, we restrict our data vector to those points at $r_p \ge 5$ Mpc/h. We do so because in addition to uncertainties in the nonlinear growth of structure, we must also account for small-scale uncertainties due to baryonic physics. Although the use of $\Upsilon$ statistics attempts to remove the impact of scales below $r_p^0$, this is not expected to be sufficient to fully account for these small-scale baryonic uncertainties, and it is hence typical to restrict fits to $E_G(r_p)$ to a slightly higher range of $r_p$ values.

\subsection{Linear galaxy bias, General Relativity}
\label{subsec:linb_GR}

We initially consider the effect of imposing Eq.~\ref{chisqreq} in the simplest (and standard in the literature) case: with $\vec{D}_{\rm nonlin}$ modelled as in GR with {\tt halofit} \citep{Takahashi2012}, and assuming a linear galaxy bias model. 
Within our LSST + DESI set up, imposing Eq.~\ref{chisqreq} results in the same `linear-only' scale cuts for both LSST Year 1 and Year 10 source samples. We attribute this perhaps-counterintuitive agreement to the fact that as seen above in Figure \ref{fig:joint_cov_corr}, the difference in shape-noise between LSST Year 1 and Year 10 is overall a fairly small effect on the covariance structure. We illustrate this in Figure \ref{fig:scales_to_keep}. We see that we are able to retain the entire $E_G(r_p)$ data vector, as compared to its constituents $\Upsilon_{gm}(r_p)$ and $\Upsilon_{gg}(r_p)$ where we only retain a small number of larger-scale data points. This is because, as seen in Eq.~\ref{egreyes}, numerator and denominator of $E_G(r_p)$ both contain a single two-point function of the matter field. Assuming (as we have done here) a linear galaxy bias, this effectively means that nonlinear behaviour in the matter field is `cancelled out'. This is both a curse and a blessing: it means that $E_G$ is unlikely to be sensitive to canonical nonlinear smoking-gun behaviour of alternative theories of gravity, like screening. On the other hand, it potentially offers a very clear way of honing in on linear-only physics in a test of modified gravity, which is a much sought-after ability.

\subsection{Alternative theories of gravity}
\label{subsec:linb_MG}

The validity of Eq.~\ref{chisqreq} as a linear-scale-cuts criterion relies on the nonlinear behaviour being actually described by GR -- or at minimum, the true nonlinear behaviour becoming important at the same scales as in GR. 
To investigate whether allowing for nonlinear modelling in other theories of gravity alters the results of the previous subsection, we repeat the application of the criterion given in Eq.~\ref{chisqreq} using modelling of $\vec{D}_{\rm lin}$ and $\vec{D}_{\rm nonlin}$ in two different alternative theories of gravity (introduced in Section \ref{subsec:mg} above):
\begin{itemize}
    \item{Hu-Sawicki $f(R)$ gravity: $f_{R0}=10^{-4}$, $n=1$.}
    \item{nDGP gravity: $\Omega_{\rm rc}=0.25$. }
\end{itemize}
We select parameter values for Hu-Sawicki $f(R)$ and nDGP gravity which are significantly discrepant from GR (in the sense that the matter power spectrum under these theories and at these parameter values deviates from GR significantly) in order to amplify any possible impact. We will take a range of values for these parameters in the next Section.

We find that when we apply Eq.~\ref{chisqreq} in both Hu-Sawicki $f(R)$ and nDGP, we retain precisely the same data points as we do in the GR case in the previous subsection. This holds across the $\Upsilon_{gm}$, $\Upsilon_{gg}$ and $E_G$ data vectors, as well as for both Year 1 and Year 10 LSST sources. Although these are only two examples of alternative gravity theories, we can cautiously state that the conclusions of Section \ref{subsec:linb_GR} -- that $E_G$ offers a possible probe of gravity that is less sensitive to nonlinear effects than its constituent probes -- appear to be robust to differences in the underlying true theory of gravity. We do, however, note that this test has not checked the impact of a theory where there is a scale-dependent difference between the lensing and Newtonian potentials.

\subsection{Nonlinear galaxy bias}
\label{subsec:scaledepbias}
The definition of $E_G$ is near-symmetric in the numerator and denominator with respect to the two-point correlation function of the matter field (modulo the lensing kernel), but it is not symmetric in galaxy bias. If scale-dependent galaxy bias is present, this effect will never cancel, as $\Upsilon_{gg}$ has two factors of this galaxy bias, $\Upsilon_{gm}$ has one, and $\beta$ will only ever include linear galaxy bias behaviour as it is definitionally measured at large scales. Measurements of $E_G$ typically must correct for any such nonlinear galaxy bias; for a general discussion of the potential effect of leaving this effect uncorrected, see \cite{leonard2015testing}.

To investigate the potential impact of this on linear scale cuts in an $E_G$ analysis, we repeat the procedure of Section \ref{subsec:linb_GR}, but this time we allow the nonlinear data vector $\vec{D}_{\rm nonlin}$ to also include nonlinear galaxy bias, while the linear data vector continues to exhibit linear galaxy bias. Nonlinear matter clustering behaviour is modelled in GR.

We are interested in the case of a nonlinear bias model which accurately reflects DESI LRGs. For this purpose, we take the Lagrangian Perturbation Theory (LPT) model with parameters as constrained in \cite{kitanidis2021cross} for a DESI-LRG-like sample from the DECam Legacy Survey (DECaLS). This work finds best-fit nonlinear bias parameters $b_1=1.333$ and $b_2=0.514$ (with $b_s$ and higher order terms fixed to 0; see e.g. \citep{chen2020consistent} for definitions of these parameters). In practice, we convert these parameters to their Eulerian equivalents values using the relations found in \cite{chen2020consistent} (Eq.~4.8) and use the {\tt FAST-PT} code \citep{mcewen2016fast} as linked through the Core Cosmology Library ({\tt CCL}, \cite{chisari2019core}) to compute clustering statistics with nonlinear biases. The equivalent Eulerian parameters are given as $b_1=2.33$, $b_2 = 1.02$, and $b_s=-0.381$ (noticing $b_s$ is non-zero in the Eulerian case). 

\begin{figure}
\centering
\includegraphics[width=0.4\textwidth]{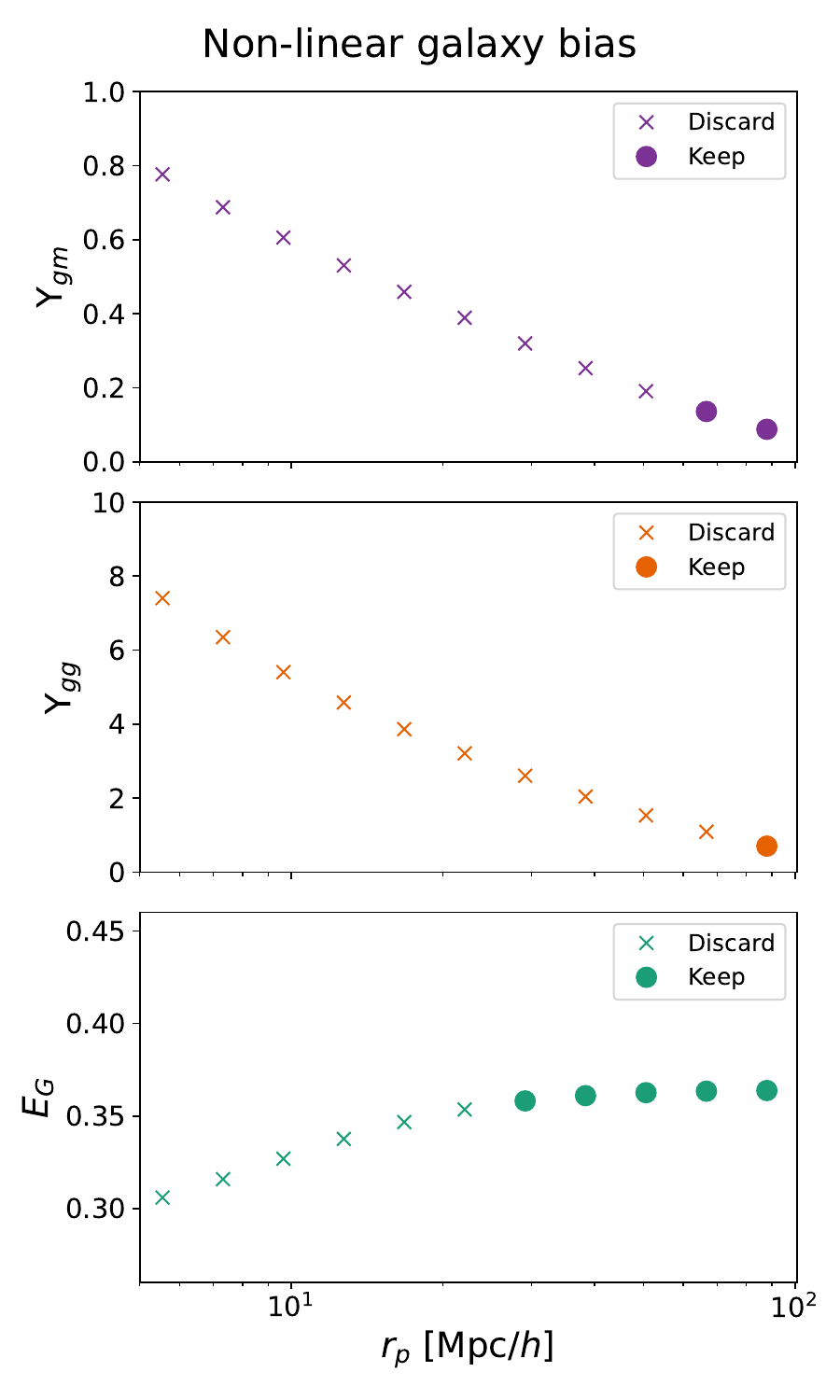}
\caption{Illustration of which points are kept in the linear-only scale cuts scenario for $\Upsilon_{gm}$ (top), $\Upsilon_{gg}$ (middle) and $E_G$ (bottom), assuming a nonlinear galaxy bias as described in Section \ref{subsec:scaledepbias}. The data vector plotted assumes this nonlinear galaxy bias model. We display the case with LSST Year 1 source galaxies; for LSST Year 10 sources, we would discard one additional point in the $E_G$ case, at $r_p \approx 30$ Mpc/h.}
\label{fig:scales_to_keep_nLbias}
\end{figure}

The effect of this nonlinear galaxy bias on which datavector elements are retained or discarded under the criterion of Eq.~\ref{chisqreq} is displayed in Figure \ref{fig:scales_to_keep_nLbias}, for LSST Year 1 sources. Note that we display the datavector under nonlinear modelling and hence we see a scale-dependence of $E_G$ as a result of nonlinear galaxy bias. For the case of LSST Year 10 sources, the data points retained and cut are very similar, with the only difference being that one additional point is cut for $E_G$ at $r_p \approx 30$ Mpc/h.

We see in Figure \ref{fig:scales_to_keep_nLbias} that the cuts to the individual $\Upsilon_{gm}$ and $\Upsilon_{gg}$ remain the same as in the linear galaxy bias case of Section \ref{subsec:linb_GR}. More pertinently, though, a large number of small-scale data points are now cut from $E_G$. As expected, this is due to $E_G$ being depressed at small scales due to the double factor of nonlinear bias in the denominator (vs the single factor in the numerator). Assuming our galaxy-bias model is reasonably close to what might be present in real DESI LRG data, it is clear that nonlinear galaxy bias must be accounted for. 

\subsection{Calibration with respect to nonlinear bias}
\label{subsec:scaledepbias_cal}

As shown in the previous subsection, nonlinear galaxy bias results, for the case of an LSST+DESI sample, in stringent linear-only scale cuts on $E_G(r_p)$. We now explore the possibility of correcting our measured $E_G(r_p)$ for this nonlinear galaxy bias, recovering a version of $E_G(r_p)$ which is insensitive to nonlinear effects at smaller scales and which could thus in principle more reliably pick up scale-dependent signatures of alternative theories of gravity.

We follow \cite{Reyes2010}, defining a similar correction factor to the one introduced in that work (for use in that case with mock galaxy catalogues):
\begin{equation}
C_b(r_p) \equiv \frac{\Omega_{\rm M}^0\Upsilon_{gg}(r_p)}{b_1 \Upsilon_{gm}(r_p)}. 
\label{eq:bias_correction}
\end{equation}
$b_1$ is the fiducial value of the linear bias parameter. $\Upsilon_{gm}(r_p)$ and $\Upsilon_{gg}(r_p)$ are computed from theory using Eqs. \ref{upgm} and \ref{upgg} respectively, in both cases crucially with nonlinear galaxy bias modelling (here, from \cite{kitanidis2021cross} as in the previous subsection). 

A corrected $E_G(r_p)$ is then defined as
\begin{equation}
\hat{E}_G(r_p) \equiv C_b(r_p)E_G(r_p).
\label{eq:corrected_Eg}
\end{equation}

To gain some intuition about this correction factor, consider that the ratio $\Upsilon_{gg}(r_p)/\Upsilon_{gm}(r_p)$ in Eq.~\ref{eq:bias_correction} is equal to $(\beta E_G(r_p))^{-1}$ (computed theoretically using the nonlinear galaxy bias model). Recall that $\beta = f / b_1$ (where $b_1$ is the linear galaxy bias parameter in the notation of this section), and that as-per Eq.~\ref{eq:eg_theory_GR}, the theoretical prediction for $E_G$ in GR is given by $f/\Omega_{\rm M}^0$. Putting these together, we see that $C_b$ is designed to capture the scale-dependent behaviour of $E_G(r_p)$ which is due to nonlinear galaxy bias, while returning to unity at large scales.  

In the case of a real measurement, $C_b$ should be calculated at the best-fit parameter values of the cosmology and nonlinear galaxy bias model. Note that a factor of $\Omega_{\rm M}^0$ will appear linearly in the theoretical prediction for $\Upsilon_{gm}(r_p)$, and as such applying the correction factor $C_b$ does not introduce additional dependence on $\Omega_{\rm M}^0$; it in indeed for this exact purpose that we include $\Omega_{\rm M}^0$ in the definition of Eq.~\ref{eq:bias_correction}.

We now go further than \cite{Reyes2010} in seeking to incorporate uncertainty associated with estimation of $C_b$ in our analysis. Particularly, we are concerned with the uncertainty which arises from the estimation of the nonlinear galaxy bias parameters.
For the current case, we refer again to the nonlinear galaxy bias model fit to a DESI-LRG-like sample, from \cite{kitanidis2021cross}. In addition to the best-fit values as given in Section \ref{subsec:scaledepbias}, that work found joint posteriors of the Lagrangian bias parameters $b_1$ and $b_2$. Taking the marginalised 2D posterior distribution over these two parameters to be approximately Gaussian (in agreement with \cite{kitanidis2021cross}), we determine a covariance matrix that reproduces their 2D marginalised constraints. From the associated 2D Gaussian joint posterior, we draw realisations of these two Lagrangian bias parameters and convert these to the associated Eulerian parameters at each realisation. 

At each set of bias parameter values, we then compute $C_b(r_p)$ using Eq.~\ref{eq:bias_correction}. From these realisations, we use the standard sample covariance relationships to compute ${\rm Cov}(C_b(r_p), C_b(r_p'))$. Note that we neglect the effect of the Hartlap correction arising from this finite number of samples when inverting the covariance as we expect it to be subdominant to that associated with the finite number of simulation realisations used in computing the covariance of the raw $E_G(r_p)$ (using the methodology of Section \ref{sec:cov}).

\begin{figure*}
\centering
\includegraphics[width=0.45\textwidth]{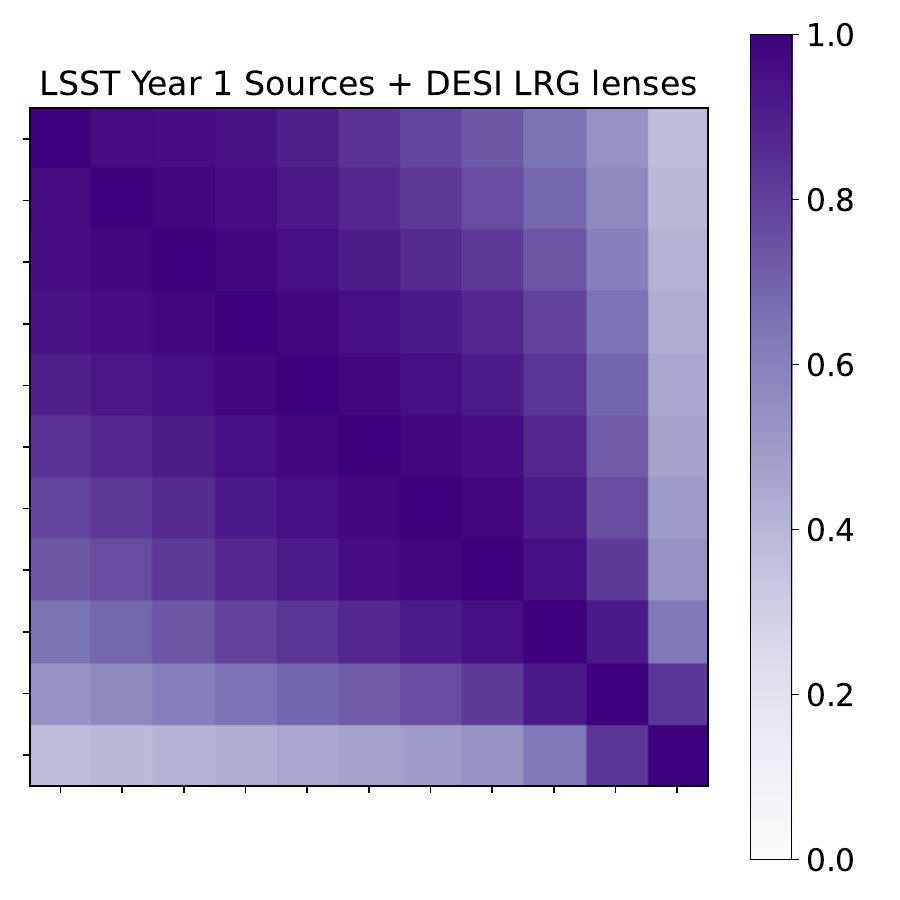}
\includegraphics[width=0.45\textwidth]{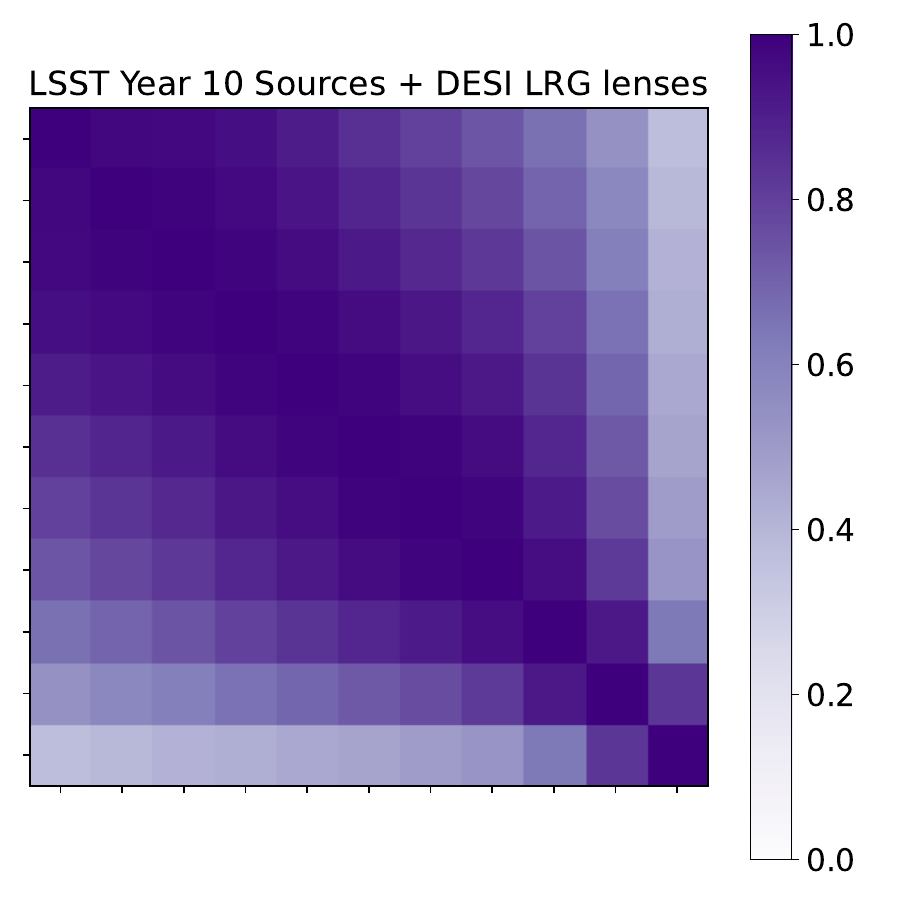}
\caption{Correlation matrices of $\hat{E}_G(r_p)$, incorporating a nonlinear bias correction factor as defined in Eqs. \ref{eq:bias_correction} and \ref{eq:cov_corrected_EG}. Bins increase logarithmically in $r_p$ with bin centres ranging from 1.4 to 87.9 Mpc/h.}
\label{fig:eg_cov_corrected}
\end{figure*}

Taking the assumption that $C_b(r_p)$ and $E_G(r_p)$ are uncorrelated, it holds that:
\begin{align}
{\rm Cov}(\hat{E}_G(r_p), &\hat{E}_G(r_p')) = \nonumber \\ &\langle C_b(r_p) \rangle \langle C_b(r_p') \rangle {\rm Cov}(E_G(r_p), E_G(r_p')) \nonumber \\ &+ \langle E_G(r_p) \rangle  \langle E_G(r_p') \rangle {\rm Cov}(C_b(r_p), C_b(r_p')).
\label{eq:cov_corrected_EG}
\end{align}
We see from the form of this equation that if the covariance of the bias correction factor is small, this reduces to the previous version of the $E_G(r_p)$ covariance modulo factors of $C_b(r_p)$. We visualise the corrected correlation matrices for the cases of both LSST Year 1 and Year 10 sources in Figure \ref{fig:eg_cov_corrected}. Comparing with Figure \ref{fig:eg_cov}, we can see that the primary effect of incorporating this correction factor on the covariance is to increase the correlation in off-diagonal elements at small $r_p$. This can be understood by recalling that nonlinear bias has an impact primarily on smaller scales.

Taking the above covariance for $\hat{E}_G(r_p)$, we compute `linear' scale cuts on the corrected $\hat{E}_G(r_p)$ using the condition of Eq.~\ref{chisqreq}, where the nonlinear data vector is now the corrected $\hat{E}_G(r_p)$.
With this correction method, we find that we are once again not required to cut any of our $\hat{E}_G(r_p)$ data points. This holds for both LSST Year 1 and LSST Year 10 sources.  We take this as our fiducial method of determining scales to be included in our analysis going forward.

It is worth noting briefly at this point that, if one's objective is to fit a model where $E_G$ is a single $r_p$-independent parameter, and to compare this value with its scale-independent GR prediction, then the discarding of smaller-scale data points may not have much effect on the ultimate error bar on this constant $E_G$ parameter. In this scenario, we are fitting a one-parameter model, and the inclusion of more data points at smaller scales does not add much information -- if (crucially) those data points are in agreement with the constant model. The benefit of including smaller scales is more obvious in the case where the data shows disagreement with an $r_p$ independent model. If this scale-dependent signature could be reliably decoupled from other scale-dependent effects (and especially nonlinear galaxy bias effects), this would be a clear indication of a deviation from GR.

\section{Testing the power of an LSST+DESI $E_G$ measurement}
\label{sec:forecast}
\noindent
We are now in a position to explore in detail the benefits and potential pitfalls of using $E_G$ to test gravity with LSST + DESI. We are interested in the question: given the true Universe is described by a theory of gravity other than GR, would we detect this?

Most literature which tests gravity using $E_G$ takes a heuristic approach to this question, qualitatively comparing measurements with the predicted value in GR at the appropriate redshift (see, e.g., \cite{Reyes2010, Blake2015, alam2017testing, jullo2019testing, blake2020testing}). This approach offers valuable intuition and has led to the strong disfavouring of some theories of gravity with extreme behaviour captured by $E_G$ (e.g. most notably Tensor-Vector-Scalar gravity (TeVeS) in \cite{Reyes2010}). However, the heuristic approach also has its pitfalls -- notably, as discussed in \cite{amon2018kids+}, the fact that the waters are invariably muddied by the fact that the GR prediction itself depends on the value of $\Omega_{\rm M}^0$. A measurement of $E_G$ which seems to disagree with the GR prediction could in fact therefore be instead indicative of the datasets in question preferring a different $\Omega_{\rm M}^0$ value than that assumed in constructing the GR prediction. 

To address this, we propose a Bayesian approach to determine whether and in what circumstances an LSST+DESI measurement of $E_G$ will detect true signatures of deviations from GR. Specifically, our approach makes use of the framework of {\it posterior predictive tests}, a Bayesian analogue to hypothesis testing \cite{gelman1996posterior} which has found considerable interest in cosmology analysis in recent years (see e.g. \cite{nicola2019consistency, porqueres2021bayesian, doux2021dark}).

\subsection{Posterior predictive test method}
\label{subsec:postpredmethod}

First, note that as in Section \ref{sec:nonlin}, we restrict to bins in $r_p \ge 5$ Mpc/h. Our method is then as follows.

{\it Generate a simulated data realisation:}
Consider a hypothetical Universe that is described by an alternative theory of gravity with specific values of the gravity-theory parameters (e.g. $f_{R0}$ in $f(R)$ gravity). We have:
\begin{itemize}
    \item{a model ($M$) for computing $E_G(r_p)$ under the alternative theory of gravity and at the specific gravity-theory parameters in question, as a function of cosmological parameters. In practice, the model is effectively insensitive to the value of any cosmological parameters other than $\Omega_{\rm M}^0$.}
    \item{a covariance matrix for $E_G(r_p)$, from Section \ref{sec:cov}. (This assumes that the covariance matrix as computed in GR is appropriate, which is a suitable approximation \cite{Kodwani2019effect}.)}
    \item{a prior on $\Omega_{\rm M}^0$, $p(\Omega_{\rm M}^0)$.}
\end{itemize}
Given these ingredients, we can generate a simulated data realisation ($D$) of $E_G(r_p)$ in this Universe. We do so by sampling from the following distribution:
\begin{equation}
    p(D|M, \Omega_{\rm M}^0)p(\Omega_{\rm M}^0).
    \label{eq:sampledata}
\end{equation}
Practically, we first draw from $p(\Omega_{\rm M}^0)$, and plug the resulting value of $\Omega_{\rm M}^0$ into the model $M$. This generates the mean of a multivariate Gaussian (with dimensionality equal to the number of $r_p$ bins) from which we draw the realisation of $D$.

{\it Fit a constant model:}
We initially fit a constant model to the data realisation drawn in the previous step, and check its goodness of fit via a $\chi^2$ test. If the fit is bad, we do not proceed, and consider GR to be rejected for this data realisation. 

{\it Find the posterior of the parameters of the GR model:} Assuming that the fit to a constant model is acceptable, we proceed to fit the parameters of the GR model for $E_G$ to the simulated data realisation. Recall that the GR model for $E_G$ is given by Eq.~\ref{eq:eg_theory_GR}, which means that the only parameter is $\Omega_{\rm M}^0$. Notationally we would like to separate this parameter $\Omega_{\rm M}^0$ from that which is used to generate the data; we call the current parameter $\Omega_{\rm M}^{0, {\rm fit}}$. In determining the posterior distribution $p(\Omega_{\rm M}^{0, {\rm fit}} | D)$, we incorporate a prior on $\Omega_{\rm M}^{0, {\rm fit}}$, which is identical to our data-generating prior on $\Omega_{\rm M}^0$ introduced above.

{\it Generate replicated data:} We draw samples from the posterior distribution on $\Omega_{\rm M}^{0, {\rm fit}}$. We use these sampled values of $\Omega_{\rm M}^{0, {\rm fit}}$ to compute corresponding instances of $E_G(r_p)$ {\it under the GR model}, $M_{\rm GR}$, which is given by Eq.~\ref{eq:eg_theory_GR} (and is constant across $r_p$ bins). This set of $E_G$ values is conventionally known as the `replicated data', $D_{\rm rep}$, within the posterior predictive framework.

{\it Compare simulated data realisation with distribution of replicated data:} We now have samples $D_{\rm rep}$, as well as the original data realisation $D$ (which, recall, is drawn from a Universe described by an alternate theory of gravity with specific parameters). We now want to determine where the original data realisation lies with respect to the distribution of replicated data. In doing so, we apply a threshold at which we consider the null GR hypothesis to be rejected, in analogy to the p-value thresholds applied in frequentist hypothesis testing. This serves only to summarise whether a given simulated analysis has accepted or rejected GR, and thus in what fraction of simulated analyses each outcome occurs. In frequentist hypothesis testing, a threshold of $p=0.05$ is typical, and we follow that standard here by considering the GR hypothesis to be rejected if the true data falls outside the symmetric 95\% probability density region of the distribution of replicated data. One could in theory apply a higher threshold to reject the null GR hypothesis, by replacing 95\% with a larger value.

There is a slight complication due to the in-principle multidimensional nature of the data (having dimension equal to the number of bins in $r_p$). The posterior predictive test requires us to map the true (simulated) data realisation as well as the samples of replicated data onto a scalar test statistic. Given that the replicated data will always be scalar-valued (due to the scale independence of $E_G$ in GR), and we have already checked that a constant model is a good fit to the data realisation, we take the constant value fit to the simulated data realisation as this test statistic. An alternative option would have been to treat the measurement of $E_G(r_p)$ in each $r_p$ bin individually from the beginning, repeating the analysis for each bin (skipping the `Fit to a constant model' step above). This would result in acceptance or rejection of GR given each data point, and would probe the question of scale-dependence indirectly. However, the power of the posterior predictive test for each individual $r_p$ bin would likely be reduced relative to the combined approach that we use here.

The above process provides the Bayesian analogue to a two-sided $p$-value, telling us whether to accept or reject GR given this specific data realisation. In an analysis using real data, we would perform the steps discussed above just once. We would thus determine where the true measured value of $E_G$ falls within the distribution of replicated data, and discuss the acceptance or rejection of the null GR hypothesis on this basis (assuming our measured $E_G(r_p)$ is consistent with scale-independence). However, in this work, we are interested in assessing the power of an LSST+DESI measurement of $E_G$. We therefore repeat this process many times for different simulated data realisations (1000 times for each scenario considered). We ultimately report the fraction of such trials for which we reject GR, for a given scenario. This demonstrates the reliability that we will detect a true underlying deviation from GR with this method in the various scenarios considered.

\subsection{Results}
\label{subsec:mg_post_nosys}

\begin{table*}
\centering
\begin{tabular}{ |c|c|c|c|c|c|c| } 
 \hline
   $\Omega_{\rm M}^0$ prior & $f_{R0}= 10^{-6}$ & $f_{R0}= 10^{-5}$ & $f_{R0}= 10^{-4}$ & $\Omega_{\rm rc} = 0.05$ & $\Omega_{\rm rc} = 0.25$ & $\Omega_{\rm rc}=0.5$ \\
\hline
 Stage III 3$\times$2pt & 0.017 & 0.018 & 0.023 & 0.017 & 0.017 & 0.017\\ 
 {\it Planck} & 0.346 & 0.572  & 0.924 & 0.345 & 0.457 & 0.559\\ 
 \hline
\end{tabular}
\caption{Fraction of cases in which we reject the GR null hypothesis for LSST Year 1 + DESI.} 
\label{tab:MG_nosys_Y1}
\centering
\end{table*}

\begin{table*}
\centering
\begin{tabular}{ |c|c|c|c|c|c|c| } 
 \hline
   $\Omega_{\rm M}^0$ prior & $f_{R0}= 10^{-6}$ & $f_{R0}= 10^{-5}$ & $f_{R0}= 10^{-4}$ & $\Omega_{\rm rc} = 0.05$ & $\Omega_{\rm rc} = 0.25$ & $\Omega_{\rm rc}=0.5$ \\
\hline
 Stage III 3$\times$2pt & 0.017 & 0.017 & 0.021 & 0.017 & 0.017 & 0.017 \\ 
  {\it Planck} & 0.319 & 0.553 & 0.925 & 0.327  & 0.438 & 0.536 \\ 
 \hline
\end{tabular}
\caption{Fraction of cases in which we reject the GR null hypothesis for LSST Year 10 + DESI.}
\label{tab:MG_nosys_Y10}
\centering
\end{table*}

\begin{figure*}
\centering
\includegraphics[width=\textwidth]{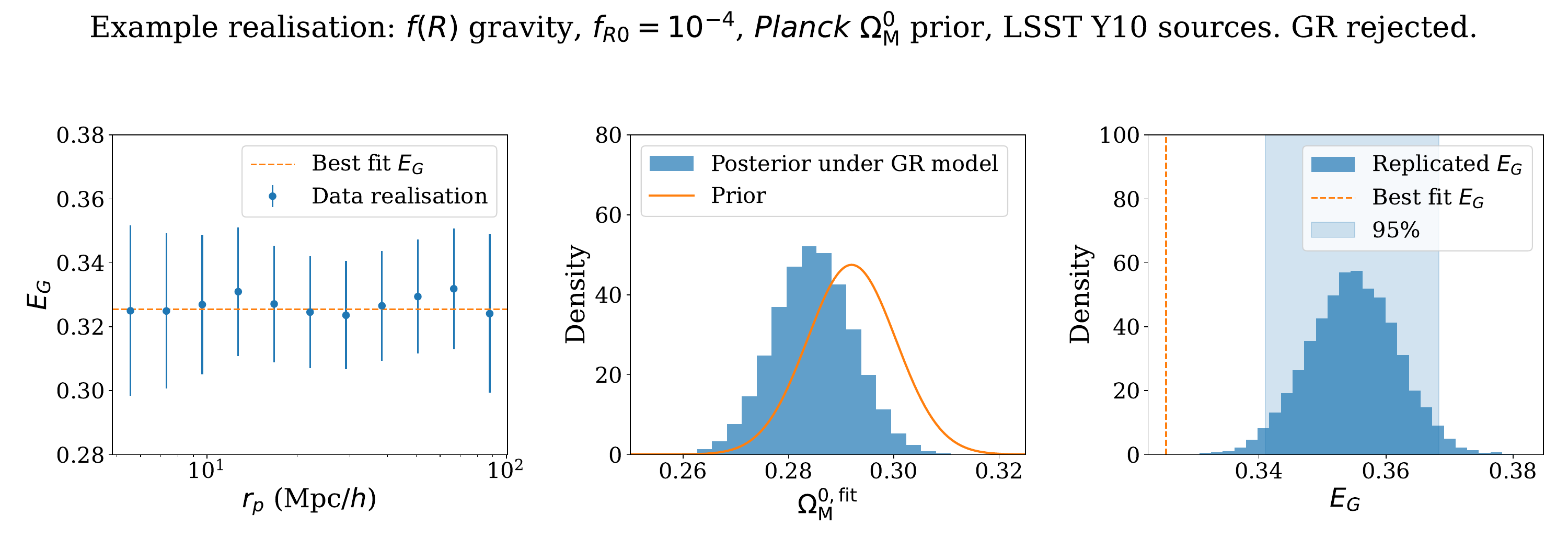}
\caption{An example where we reject the GR null hypothesis when $f(R)$ gravity is the underlying true gravity theory. We show an example data realisation and the resulting steps of the posterior predictive method for the case of $f_{R0}=10^{-4}$, LSST Year 10 sources, and a {\it Planck} CMB prior on $\Omega_{\rm M}^0$. The left panel shows the data realisation and best fit $E_G$ constant value. The middle panel is the posterior distribution on $\Omega_{\rm M}^{0, {\rm fit}}$. The rightmost panels is the resulting replicated $E_G$ values (`$D_{\rm rep}$'), compared with the best fit value of $E_G$ with respect to the data realisation and covariance. We see that in this example, we reject the GR null hypothesis.}
\label{fig:example_reject_fR}
\end{figure*}

\begin{figure*}
\centering
\includegraphics[width=\textwidth]{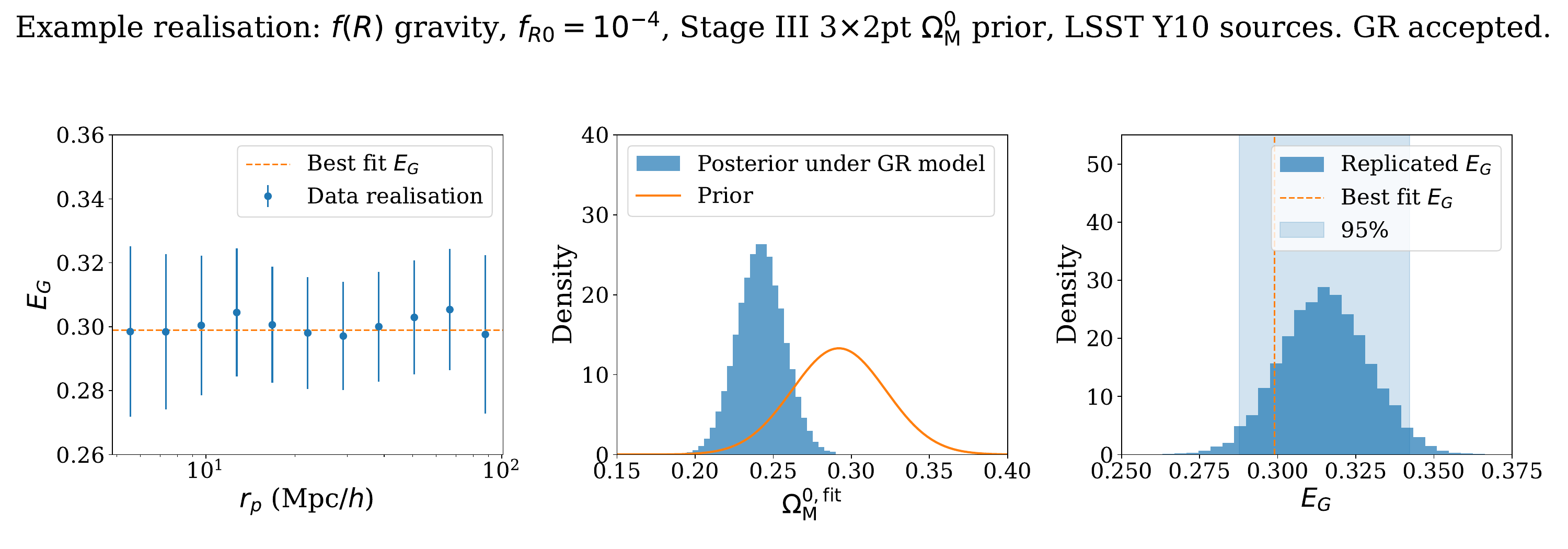}
\caption{An example where we accept the GR null hypothesis when $f(R)$ gravity is the underlying true gravity theory. We show an example data realisation and the resulting steps of the posterior predictive method for the case of $f_{R0}=10^{-4}$, LSST Year 10 sources, and a Stage III $3\times2$pt (Dark Energy Survey Year 3) prior on $\Omega_{\rm M}^0$. The left panel shows the data realisation and best fit $E_G$ constant value. The middle panel is the posterior distribution on $\Omega_{\rm M}^{0, {\rm fit}}$. The rightmost panels is the resulting replicated $E_G$ values (`$D_{\rm rep}$'), compared with the best fit value of $E_G$ with respect to the data realisation and covariance. We see that in this example, we accept the GR null hypothesis.}
\label{fig:example_accept_fR}
\end{figure*}

\begin{figure*}
\centering
\includegraphics[width=\textwidth]{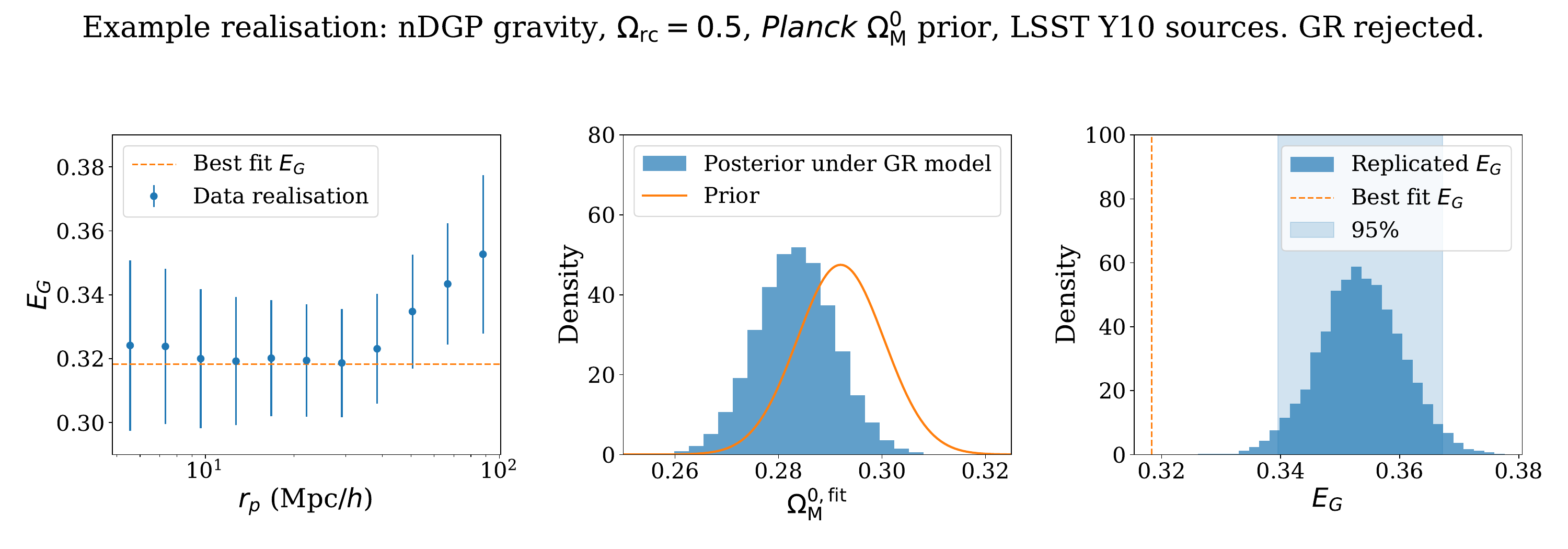}
\caption{An example where we reject the GR null hypothesis when nDGP gravity is the underlying true gravity theory. We show an example data realisation and the resulting steps of the posterior predictive method for the case of $\Omega_{\rm rc}=0.5$, LSST Year 10 sources, and a {\it Planck} CMB prior on $\Omega_{\rm M}^0$. The left panel shows the data realisation and best fit $E_G$ constant value. The middle panel is the posterior distribution on $\Omega_{\rm M}^{0, {\rm fit}}$. The rightmost panels is the resulting replicated $E_G$ values (`$D_{\rm rep}$'), compared with the best fit value of $E_G$ with respect to the data realisation and covariance. We see that in this example, we reject the GR null hypothesis.}
\label{fig:example_reject_nDGP}
\end{figure*}

\begin{figure*}
\centering
\includegraphics[width=\textwidth]{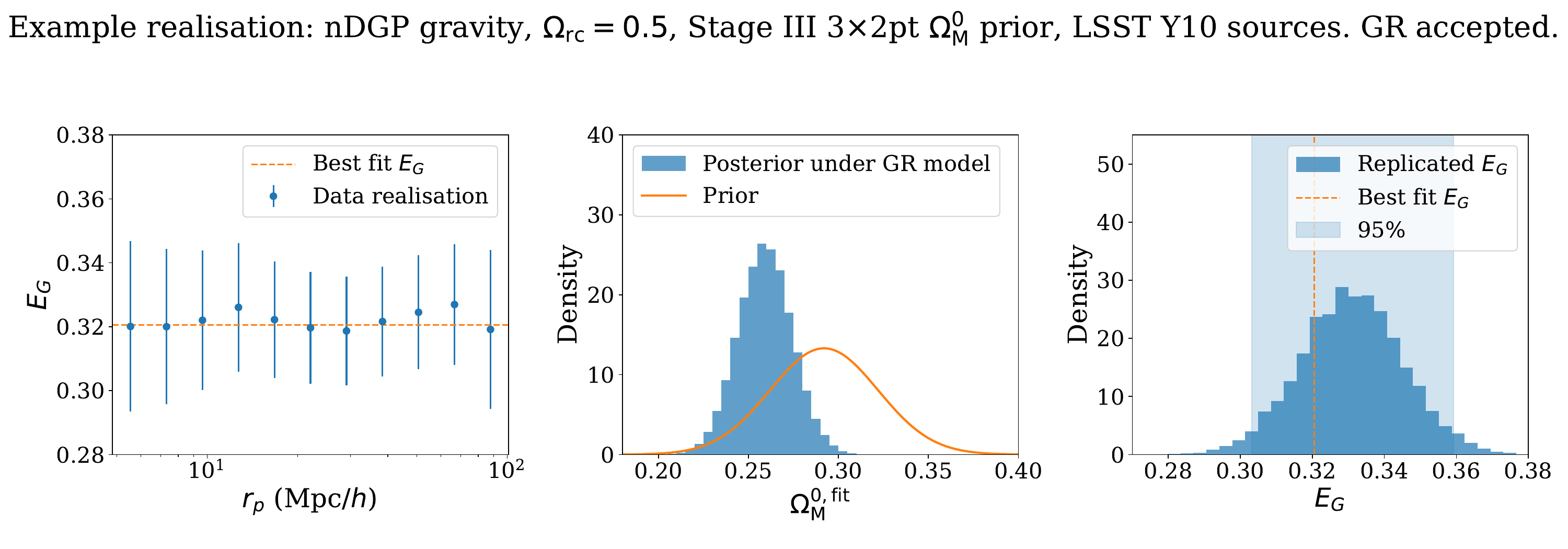}
\caption{An example where we accept the GR null hypothesis when nDGP gravity is the underlying true gravity theory. We show an example data realisation and the resulting steps of the posterior predictive method for the case of $\Omega_{\rm rc}=0.5$, LSST Year 10 sources, and a Stage III $3\times2$pt (Dark Energy Survey Year 3) prior on $\Omega_{\rm M}^0$. The left panel shows the data realisation and best fit $E_G$ constant value. The middle panel is the posterior distribution on $\Omega_{\rm M}^{0, {\rm fit}}$. The rightmost panels is the resulting replicated $E_G$ values (`$D_{\rm rep}$'), compared with the best fit value of $E_G$ with respect to the data realisation and covariance. We see that in this example, we accept the GR null hypothesis.}
\label{fig:example_accept_nDGP}
\end{figure*}

We will consider the following cases for the underlying `true' alternative gravity models:
\begin{itemize}
    \item{nDGP gravity with $\Omega_{\rm rc} = \{0.05, 0.25, 0.5\}$}
    \item{Hu-Sawicki $f(R)$ gravity with $f_{R0}=\{10^{-6}, 10^{-5}, 10^{-4}\}$ and $n=1$}
\end{itemize}

We consider two options for the prior on $\Omega_{\rm M}^0$. In both cases, we use a Gaussian prior with mean shifted to agree with our fiducial parameters; the priors differ only by their variances.
\begin{itemize}
    \item{Stage III 3$\times$2pt prior: $\sigma(\Omega_{\rm M}^0)= 0.03$, corresponding to the constraint on $\Omega_{\rm M}^0$ under the $\Lambda$CDM model from Dark Energy Survey (DES) Year 3 \cite{DESY3Key}.}
    \item{CMB prior: $\sigma(\Omega_{\rm M}^0)= 0.0084$, corresponding to the constraint on $\Omega_{\rm M}^0$ under the $\Lambda$CDM model (temperature and E-mode polarization auto- and cross-spectra case case) from {\it Planck} \cite{Planck2018}.} 
\end{itemize}

We find that in the case of the 3$\times$2pt prior on $\Omega_{\rm M}^0$, for both LSST Year 1 and Year 10 sources, we have a very low fractional rate of rejecting the GR null hypothesis in the presence of true gravity described by any of our alternate models. However, when we move to the more restrictive CMB prior on $\Omega_{\rm M}^0$, the rate at which we reject the null hypothesis is considerably higher, particularly in the cases of true alternative gravity descriptions which deviate more from GR.  Fractional rejection rates are given in Tables \ref{tab:MG_nosys_Y1} and \ref{tab:MG_nosys_Y10} for LSST Year 1 and Year 10 respectively. We see that, as discussed qualitatively in \cite{amon2018kids+}, the prior information available on $\Omega_{\rm M}^0$ is crucial in determining the power of $E_G$ as diagnostic for modified gravity. The introduction of the tighter prior on $\Omega_{\rm M}^0$ from CMB data leads to significantly higher rejection rates of GR (recalling that in these simulated analyses, GR is never the true underlying theory of gravity). The case of the CMB prior on $\Omega_{\rm M}^0$ combined with a true gravity theory of $f(R)$ with $f_{R0}=10^{-4}$ is particularly dramatic, with a Stage III 3$\times$2pt prior offering only approximately 2\% rejection of GR, whereas the CMB prior offer over 92\% rejection rate in both LSST Year 1 and Year 10. Furthermore, the effect of the tighter vs wider $\Omega_{\rm M}^0$ prior far outweighs the effect of moving from LSST Year 1 to Year 10 sources.

To better understand what is happening, we consider in Figures \ref{fig:example_reject_fR} and \ref{fig:example_accept_fR} two example data realisations under two different cases, and illustrate the steps of the posterior predictive method in each case. In both scenarios, the underlying true theory of gravity is $f(R)$ gravity with $f_{\rm R0}=10^{-4}$, and the sources are from LSST Year 10. In Figure \ref{fig:example_reject_fR}, the prior on $\Omega_{\rm M}^0$ is from {\it Planck}, and the null hypothesis of GR is ultimately rejected, whereas in Figure \ref{fig:example_accept_fR}, the prior is from DES Y3, and the null hypothesis is accepted. Recall first that both priors are Gaussians centred on the fiducial value of $\Omega_{\rm M}^0$ (here, 0.292) differing only in their variances. In examining particularly the middle panel of both figures, we see that in both cases, the non-GR data is pulling the posterior towards lower values of $\Omega_{\rm M}^{0, {\rm fit}}$. However, in Figure \ref{fig:example_accept_fR}, the less informative prior has less impact in pulling the posterior up to a higher value of $\Omega_{\rm M}^{0, {\rm fit}}$ than does the more informative prior in Figure \ref{fig:example_reject_fR}. The result is that in the case shown in Figure \ref{fig:example_accept_fR} we accept GR, with the implicit corollary that we are allowed to be in a Universe with $\Omega_{\rm M}^0$ described by the middle panel of Figure \ref{fig:example_accept_fR}. On the other hand, in the case of a tighter {\it Planck} CMB prior as in Figure \ref{fig:example_reject_fR}, the prior drags the posterior on $\Omega_{\rm M}^{0, {\rm fit}}$ closer to the fiducial value, with correspondingly higher $E_G$ values in the replicated data distribution (rightmost panel), which are inconsistent with the best fit value to the $E_G$ data realisation in the non-GR Universe. As a result, it is this more strict prior information on $\Omega_{\rm M}^0$ that leads us to (correctly) reject the null GR hypothesis in this case. For completeness, we present equivalent cases under nDGP gravity in Figures \ref{fig:example_reject_nDGP} and \ref{fig:example_accept_nDGP}.

Zooming back out from these specific examples to the broad picture: for scenarios with the Stage III $3\times$2pt prior on $\Omega_{\rm M}^0$, the strong majoirty of cases of rejecting GR are due to the constant model being flagged as a bad fit by the reduced $\chi^2$ test, rather than due to the posterior predictive test. Because the actual theoretical predictions from both $f(R)$ and nDGP are scale-independent, we attribute these failures to fit a constant model instead to fluctuations in the data-generating draws. This is in contrast to the case of the CMB prior, where rejecting GR is nearly always due to the posterior predictive test.

We note also that even in the case of the CMB prior, differences between the rate of null hypothesis rejection in the case of LSST Year 1 vs Year 10 sources are minimal. In fact, we see that the LSST Year 10 case actually rejects GR slightly less frequently than for LSST Year 1 (although these differences minor). We therefore conclude that the dominant effect for this data set-up is that of the prior on $\Omega_{\rm M}^0$, rather than the shape noise of the source sample.

Finally, it is worth briefly noting that the priors we have considered on $\Omega_{\rm M}^0$ are taken from analyses performed under the assumption of GR. Analyses under the assumption that a different theory of gravity holds would most likely have resulted in a broader constraints on $\Omega_{\rm M}^0$, due to their increased parameter space (see e.g. \cite{hu2016testing, wang2021can, piga2023constraints}). For real-data analyses, a conservative approach that takes the $\Omega_{\rm M}^0$ prior from analyses that allow for additional degrees of freedom in the gravitational sector should be considered.

\section{Discussion and Conclusions}
\label{sec:conc}
\noindent

In this work, we have explored the potential of testing gravity with $E_G$ using LSST weak lensing source galaxies and DESI LRG lenses. We construct the required covariance matrices and examine the sensitivity of $E_G$ to nonlinear effects in structure formation and galaxy bias. We have introduced the idea of the posterior predictive test as a valuable tool for consistently incorporating theoretical uncertainties in $E_G$ (from its model-dependence on $\Omega_{\rm M}^0$).

Our work highlights in a quantitative manner the key role played by the prior information available on $\Omega_{\rm M}^0$ in the power of an LSST+DESI $E_G$ measurement to reject the null GR hypothesis in the presence of non-GR gravitational behaviour. We have seen that the effect of imposing a {\it Planck} prior on $\Omega_{\rm M}^0$ as opposed to a Stage III 3$\times$2pt prior has a much more significant impact on our capacity to reject the null GR hypothesis than moving from an LSST Year 1 to an LSST Year 10 source sample, for the alternative theories of gravity considered here. This result highlights that which has been suggested in the literature previously \cite{amon2018kids+}: that we are now well outside the regime where the uncertainty on $\Omega_{\rm M}^0$ can be disregarded when interpreting $E_G$ measurements. We therefore strongly advocate the use of an analysis methodology which consistently incorporates this factor, such as the posterior predictive test as we have used here. 

In the case of the stronger CMB priors on $\Omega_{\rm M}^0$, we reject GR in the strong majority of cases only for Hu-Sawicki gravity with $f_{R0}=10^{-4}$ (out of the scenarios considered). As this theory and parameter combination is generally considered in the literature to be an unrealistically large deviation from GR, this is perhaps not terribly promising at face value. However, we emphasise that the theories under consideration here were selected not because they are considered representative of the most likely ways nature may deviate from General Relativity. Hu-Sawicki $f(R)$ and nDGP are workhorse examples of alternative gravity theories. We selected these in part because they cover two major screening mechanisms prevalent amongst alternative theories of gravity, but also because of the availability of tools for computing cosmological observables in these theories. The apparent pessimism of our results should therefore be taken with a grain of salt, particularly in combination with our major finding, which is the extreme role of prior information on $\Omega_{\rm M}^0$ in determining our ability to reject the null GR hypothesis.

One further caveat to the selection of gravity theories considered here is that in neither of these theories do we expect a scale-dependent $E_G(r_p)$. To be clear, this scenario does capture many alternative gravity theories of interest. In all theories with a single scalar degree of freedom, for example, the linear modification to the growth and lensing potential exhibit scale-dependence only sub-dominantly, becoming relevant only at near-horizon scales which are not relevant to us here \citep{BakerScales2014}. In the original paper proposing $E_G$ \citep{Zhang2007}, a single theory (TeVeS \cite{skordis2009tensor}) was cited as being expected to exhibit a scale-dependent $E_G(r_p)$; this was strongly disfavoured by the first measurement of $E_G$ \citep{Reyes2010} not long after. However, for completeness we highlight that our results may not hold in the case where the true Universe is described by a theory which would induce strong scale-dependence in $E_G$; presumably, in such a case, an LSST+DESI measurement $E_G$ would be more sensitive to deviations from GR than we have seen here.

In our exploration of the sensitivity of $E_G$ to nonlinear behaviour (Section \ref{sec:nonlin}), we found that if we assume a decoupling between the scale-dependence of nonlinear structure formation at a dark matter only level, and that of galaxy bias (not a trivial assumption), $E_G$ may offer a method for testing gravity uniquely insensitive to uncertainties in nonlinear structure formation. Although this means that $E_G$ is unlikely to be sensitive to screening effects, it does open the door for the possibility of different ways of utilising this combination of measurements - perhaps in combination with linear-only parameterisations of deviations from GR as a data vector used to constrain parameters.

While we have presented in this work a thorough study of the effect of nonlinear structure growth and galaxy bias on $E_G$ for LSST+DESI, this is not an exhaustive consideration of potentially relevant systematic effects. Particularly, future work should consider to what extent un-modelled galaxy intrinsic alignment effects in $\Upsilon_{gm}$ could induce spurious detections of deviations from GR with $E_G$. This was considered in  \citep{rauhut2025testing} for DESI Data Release 1 LRG lenses and Stage III weak lensing sources, where it was found that accounting for intrinsic alignment has a very minor impact on results, however, this may not carry over to the case of LSST sources. Similarly, although we expect $E_G$ to be largely insensitive to the details of the source galaxy redshift distribution, verifying this explicitly for realistic LSST photometric redshift uncertainties would be valuable. Should either intrinsic alignment or photometric redshift uncertainties prove to be important systematic effects for $E_G$, it is likely that careful selection of lens and source tomographic bins (minimizing the number of lens-source pairs which are close along the line of sight) could mitigate their impacts.

LSST promises to deliver an unprecedented wealth of high-quality data, making it a uniquely powerful tool for testing gravity on cosmological scales. Consistency-check analyses, of which $E_G$ is a major example, uniquely offer a way of testing gravity that does not require the assumption of a parametrization or particular theory of interest within our analysis. Our results provide a clear indication that for $E_G$ to play a powerful role as a consistency check with LSST and DESI data requires precision prior knowledge on $\Omega_{\rm M}^0$. 

\begin{acknowledgments}
This paper has undergone internal review in the LSST Dark Energy Science Collaboration. 
The internal reviewers were Jessie Muir and Luisa Jaime. 
Author contributions: CDL wrote code, performed the main analysis, wrote most of the paper text, and drove the project. SA wrote code and performed analysis to enable the calculation of simulation-based covariance matrices. RM participated in formulating the project idea, provided scientific guidance, and participated in paper revision. MMR participated in formulating the posterior predictive methodology. SS wrote code to compute the analytic covariance matrices and provided scientific input. CZ wrote code to compute the power spectrum under alternative theories of gravity, wrote the section of the paper describing alternative gravity theories and participated in paper revision.
CDL is supported by the Science and Technology Facilities Council (STFC) [grant No. UKRI1172].
RM is supported by the Department of Energy Cosmic Frontier program, grant DE-SC0010118.
The DESC acknowledges ongoing support from the Institut National de 
Physique Nucl\'eaire et de Physique des Particules in France; the 
Science \& Technology Facilities Council in the United Kingdom; and the
Department of Energy and the LSST Discovery Alliance
in the United States.  DESC uses resources of the IN2P3 
Computing Center (CC-IN2P3--Lyon/Villeurbanne - France) funded by the 
Centre National de la Recherche Scientifique; the National Energy 
Research Scientific Computing Center, a DOE Office of Science User 
Facility supported by the Office of Science of the U.S.\ Department of
Energy under Contract No.\ DE-AC02-05CH11231; STFC DiRAC HPC Facilities, 
funded by UK BEIS National E-infrastructure capital grants; and the UK 
particle physics grid, supported by the GridPP Collaboration.  This 
work was performed in part under DOE Contract DE-AC02-76SF00515.
We acknowledge the use of NumPy \cite{harris2020array}, SciPy \cite{2020SciPy-NMeth} and Matplotlib \cite{Hunter:2007}.

The analysis code for this paper is publicly available at \footnote{https://github.com/c-d-leonard/EG\_forecasts}. We provide this code publicly in the spirit of open science and make no guarantees of code maintenance.

\end{acknowledgments}

\appendix

\section{Additional covariance-related expressions and figures}
\label{app:covmaths}

In this appendix, we provide for completeness some additional expressions and figures for helpful quantities in the calculation of covariance discussed in Section \ref{sec:cov}.

The average in $r_p$ and $r_p'$ bins of Cov[$\Delta \Sigma_{gx}(r_p)$, $\Delta \Sigma_{gy}(r_p')$] (where $x$ and $y$ are given by either $g$ or $m$) is given by:
\begin{align}
{\rm Cov}&\left[\Delta \Sigma_{gx}\left(r_p^i\right),\Delta \Sigma_{gy}\left(r_p^j\right) \right] \nonumber \\&=\frac{4}{\left(\left(r_p^{i,h}\right)^2-\left(r_p^{i,l}\right)^2 \right)\left(\left(r_p^{j,h}\right)^2-\left(r_p^{j,l}\right)^2\right)}  \nonumber \\ &\times \int_{r_p^{i,l}}^{r_p^{i,h}} dr_p \, r_p \int_{r_p^{j,l}}^{r_p^{j,h}} dr_p^\prime \, r_p^\prime \, {\rm Cov}\left[\Delta \Sigma_{gx}\left(r_p\right),\Delta \Sigma_{gy}\left(r_p^\prime \right) \right]
\label{DeltaSigmaCov_avg_gen}
\end{align}
where $r_p^{i,h}$ and $r_p^{i,l}$ represent the high and low edges of bin $r_p^i$ respectively (and similarly for bin $r_p^j$). The equivalent averaged quantity for Cov[$\Upsilon_{gx}(r_p)$, $\Upsilon_{gy}(r_p')$] is then:
\begin{align}
{\rm Cov}&(\Upsilon_{gx}(r_p^i),\Upsilon_{gy}(r_p^j)) = {\rm Cov}(\Delta \Sigma_{gx}(r_p^i),\Delta \Sigma_{gy}(r_p^j))\nonumber \\ &-\frac{2 (r_p^0)^2}{(r_p^{i,h})^2 - (r_p^{i_l})^2)} \ln\left(\frac{r_p^{i,h}}{r_p^{i,l}}\right) {\rm Cov}(\Delta \Sigma_{gx}(r_p^{0}),\Delta \Sigma_{gy}(r_p^j))\nonumber \\&- \frac{2 (r_p^0)^2}{(r_p^{j,h})^2 - (r_p^{j_l})^2)} \ln\left(\frac{r_p^{j,h}}{r_p^{j,l}}\right){\rm Cov}(\Delta \Sigma_{gx}(r_p^i),\Delta \Sigma_{gy}(r_p^{0})) \nonumber \\ &+\frac{2 (r_p^0)^2}{(r_p^{i,h})^2 - (r_p^{i_l})^2)} \ln\left(\frac{r_p^{i,h}}{r_p^{i,l}}\right)\frac{2 (r_p^0)^2}{(r_p^{j,h})^2 - (r_p^{j_l})^2)} \ln\left(\frac{r_p^{j,h}}{r_p^{j,l}}\right)\nonumber \\ &\times{\rm Cov}(\Delta \Sigma_{gx}(r_p^{0}),\Delta \Sigma_{gy}(r_p^{0})) 
\label{covupgm_avg}
\end{align}

\begin{figure*}
\centering
\includegraphics[width=0.45\textwidth]{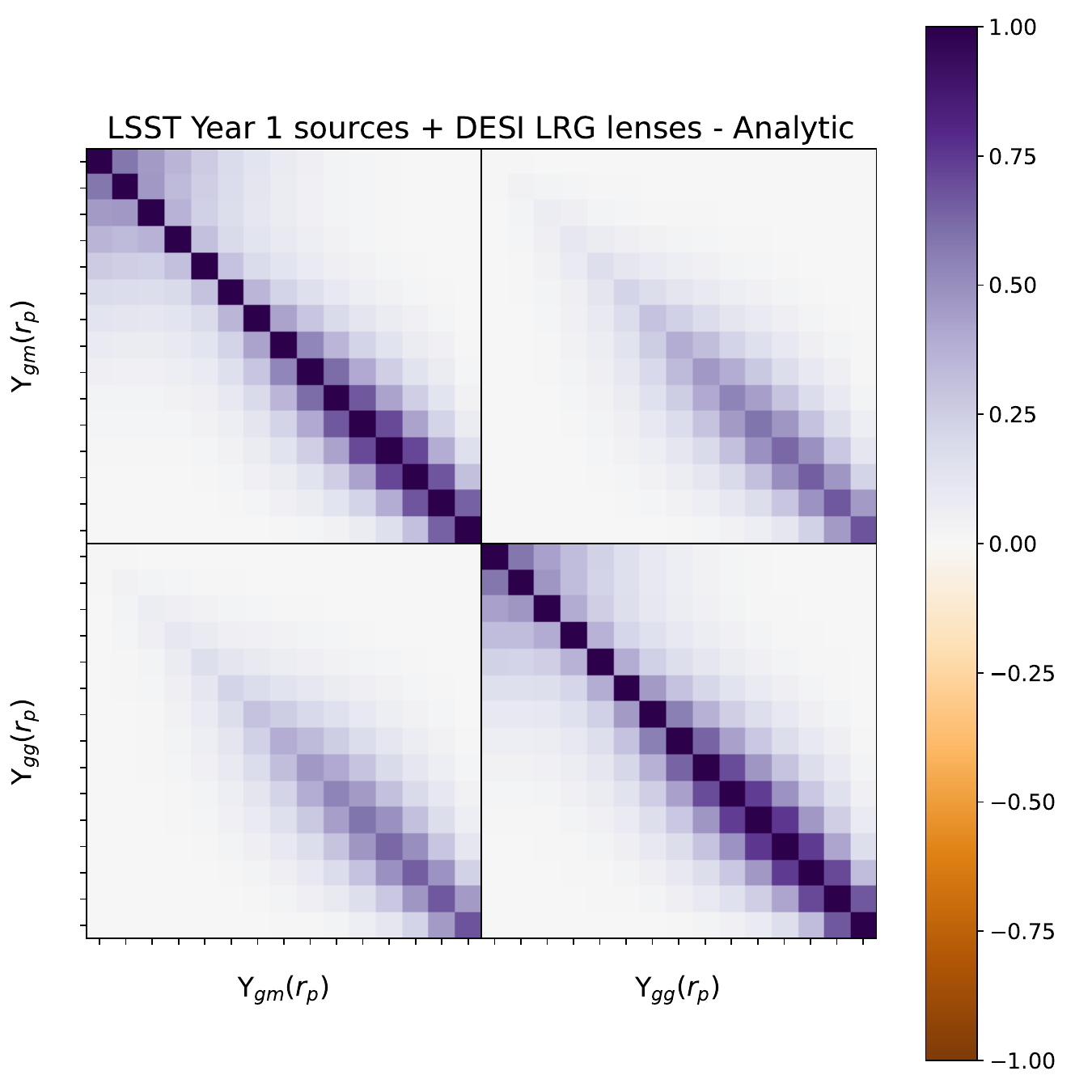}
\includegraphics[width=0.45\textwidth]{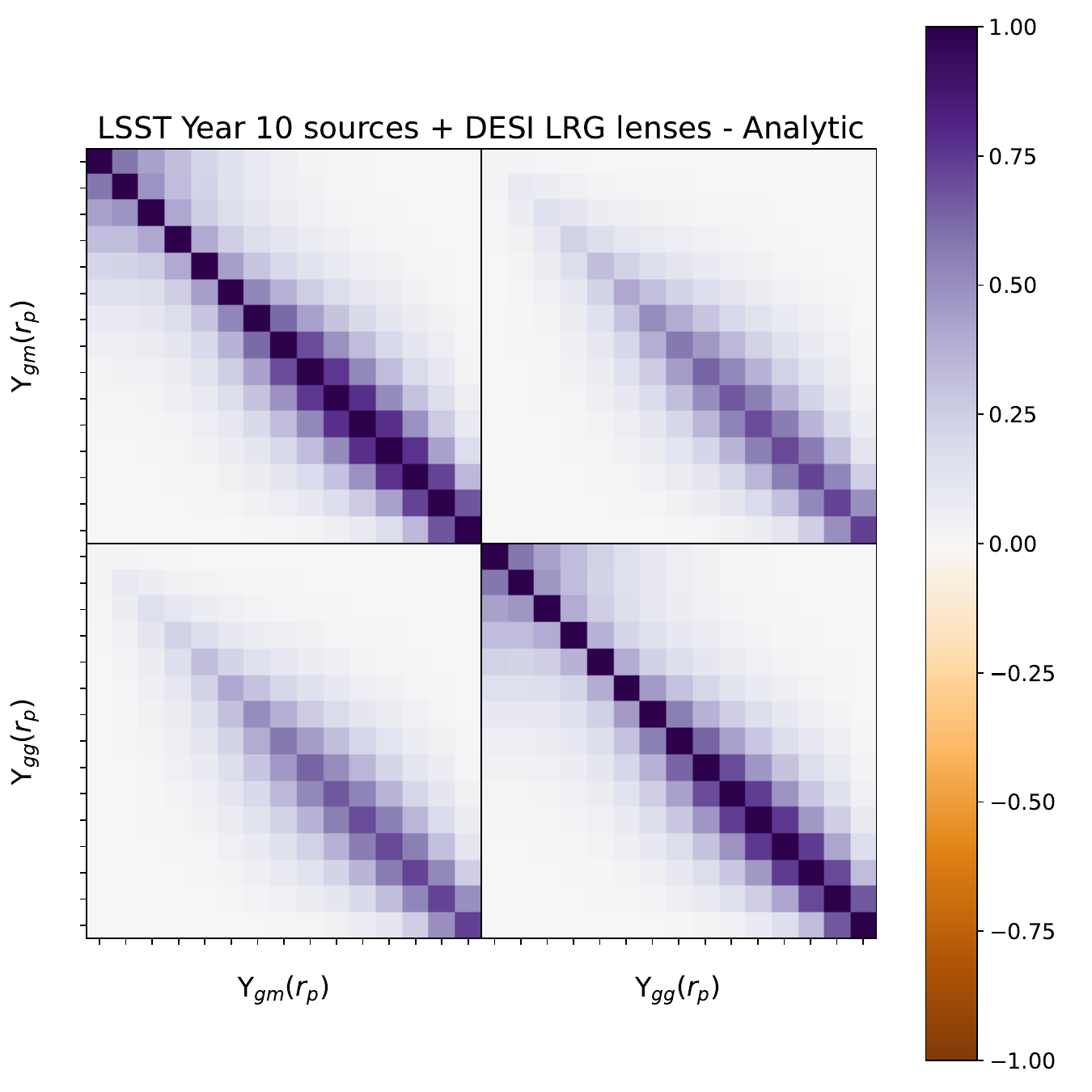}
\caption{Correlation matrix of $\Upsilon_{gm}$ and $\Upsilon_{gg}$ for LSST Year 1 (left) and Year 10 (right) sources with DESI LRG lenses, estimated using the analytic method of Section \ref{subsec:cov_elem}.}
\label{fig:joint_cov_corr_ana}
\end{figure*}

Specifically, in the case of considering the shape-noise-only term of ${\rm Cov}(\Upsilon_{gm}(r_p^i),\Upsilon_{gm}(r_p^j))$ for the purpose of combining with a simulation-estimated covariance that excludes this effect, we have:
\begin{widetext}
\begin{align}
{\rm Cov}^{\rm SN} (\Upsilon_{gm}(r_p^i),\Upsilon_{gm}(r_p^j)) &= \frac{\left(\overline{\Sigma_c^{-2}}\right)^{-1}\sigma_\gamma^2}{2\pi V n_l n_s} \Bigg(\frac{2}{\left(r_p^{i,h}\right)^2 - \left(r_p^{i,l}\right)^2} \nonumber \\
&-\frac{4 \left(r_p^0\right)^2}{\left(\left(r_p^{i,h}\right)^2 - \left(r_p^{i,l}\right)^2\right)\left(\left(r_p^{0,h}\right)^2 - \left(r_p^{0,l}\right)^2\right)} \ln\left(\frac{r_p^{i,h}}{r_p^{i,l}}\right)\delta^K_{i,0}\nonumber \\
&-\frac{4 \left(r_p^0\right)^2}{\left(\left(r_p^{j,h}\right)^2 - \left(r_p^{j,l}\right)^2\right)\left(\left(r_p^{0,h}\right)^2 - \left(r_p^{0,l}\right)^2\right)}  \ln\left(\frac{r_p^{j,h}}{r_p^{j,l}}\right)\delta^K_{j,0} \nonumber \\
&+\frac{8 \left(r_p^0\right)^4}{\left(\left(r_p^{i,h}\right)^2 - \left(r_p^{i,l}\right)^2\right)\left(\left(r_p^{j,h}\right)^2 - \left(r_p^{j,l}\right)^2\right)\left(\left(r_p^{0,h}\right)^2 - \left(r_p^{0,l}\right)^2\right)} \ln\left(\frac{r_p^{i,h}}{r_p^{i,l}}\right) \ln\left(\frac{r_p^{j,h}}{r_p^{j,l}}\right) \Bigg)
\label{covSN_avg}
\end{align}
\end{widetext}
where $\delta^K_{i,j}$ is a Kronecker delta on $i,\,j$.
The analytic correlation matrices between $\Upsilon_{gm}(r_p)$ and $\Upsilon_{gg}(r_p)$ (auto and cross) as calculated using the expressions in Section \ref{sec:cov} is displayed in Figure \ref{fig:joint_cov_corr_ana} for LSST Year 1 and Year 10 sources.

%

\end{document}